\documentclass[iop]{emulateapj}
\usepackage{color}

\shorttitle{Intrinsic alignment}
\shortauthors{Hao et al.}

\begin{document}

\title{Intrinsic Alignment of Cluster Galaxies: the Redshift Evolution}
\author{Jiangang Hao\altaffilmark{1}, Jeffrey M. Kubo\altaffilmark{1}, Robert Feldmann\altaffilmark{1,2}, James Annis\altaffilmark{1}, David E. Johnston\altaffilmark{1}, Huan Lin\altaffilmark{1}, Timothy A. McKay\altaffilmark{3,4}}

\altaffiltext{1}{Center for Particle Astrophysics, Fermi National Accelerator Laboratory, Batavia, IL 60510}
\altaffiltext{2}{Kavli Institute for Cosmological Physics, The University of Chicago, Chicago, IL 60637}
\altaffiltext{3}{Department of Physics, University of Michigan, Ann Arbor, MI 48109}
\altaffiltext{4}{Department of Astronomy, University of Michigan, Ann Arbor, MI 48109}

\begin{abstract}
We present measurements of two types of cluster galaxy alignments based on a volume limited and highly pure ($\ge$ 90\%) sample of clusters from the GMBCG catalog derived from SDSS DR7. We detect a clear BCG alignment (the alignment of major axis of the BCG toward the distribution of cluster satellite galaxies). We find that the BCG alignment signal becomes stronger as the redshift and BCG absolute magnitude decrease, and becomes weaker as BCG stellar mass decreases. No dependence of the BCG alignment on cluster richness is found. We can detect a statistically significant ($\ge$ 3 sigma) satellite alignment (the alignment of the major axes of the cluster satellite galaxies toward the BCG) only when we use the isophotal fit position angles (PAs, hereafter), and the satellite alignment depends on the apparent magnitudes rather than the absolute magnitudes of the BCGs. This suggests the detected satellite alignment based on isophotoal PAs from the SDSS pipeline is possibly due to the contamination from the diffuse light of nearby BCGs. We caution that this should not be simply interpreted as non-existence of the satellite alignment, but rather that we cannot detect them with our current photometric SDSS data. We perform our measurements on both SDSS $r$ band and $i$ band data, but did not observe a passband dependence of the alignments.  
\end{abstract}

\section{Introduction}

Galaxy orientations contain important information about the gravitational environment in which it resides. \citet{brown38} pointed out that galaxy orientations may not be isotropic due to the large scale gravitational interaction. \citet{hawley75} reported a weak evidence of anisotropy of the galaxy orientation. Galaxy orientation becomes especially interesting in the vicinity of galaxy clusters, where the strong gravitational field may produce detectable orientation preference for both central galaxies and satellite galaxies. With the advent of modern sky surveys, such as Sloan Digital Sky Survey (SDSS)~\citep{york00}, the shape and orientation of galaxies can be measured to high precision. The large sky coverage substantially increases the sample size to allow statistically significant measurements on cluster galaxy alignments, from which our understanding of the cluster formation process can be greatly improved.

The term galaxy alignment has been used extensively in the literature and refers to alignments in different contexts. In the galaxy cluster environment, there are two types of alignments that are of great interest. The first is the alignment of the major axes of the cluster satellite galaxies towards the cluster center, which we will call ``satellite alignment''. In our case, as an operational definition, we will consider the alignment between the major axes of the satellite galaxies and the BCG\footnote{Though cluster center is well defined as the deepest gravitational potential well, its determination from observational data, especially optical data, is not unambiguous. The central galaxy in a cluster (the one which resides near the bottom of the cluster potential well) is very often the brightest galaxy (BCG) in the cluster. This BCG is then coincident with the region with the deepest potential traditionally identified in theory as the center of a cluster. Using the BCG as the cluster center simplifies precise comparisons between observations and theory.}. The second type of alignment is the alignment of the BCGs major axis towards the distribution of satellite galaxies in the cluster. We will call this alignment ``BCG alignment'' hereafter. 

In addition to these two types of alignments, the possible alignments between the major axises of the satallite galaxies and the cluster~\citep{plinois04}, and between the cluster shape and large scale structures~\citep{paz08,flatenbacher09,wangyougang09,paz11} have been studied, but these are beyond the scope of this current paper.

There are extensive studies on these two types of alignments with both simulations and observations. It is argued, based on simulations, that the preferred accretion direction of satellite halos toward the host halo along the filaments is largely responsible for the BCG alignment~\citep{tormen97,vitvitska02,knebe04,zentner05,wang05}. The detections of this alignment from real data has been reported by many teams~\citep{sastry68,austin74,dressler78,carter80,binggeli82,brainerd05,yang06,azzaro07,wangyougang08,siverd09,ostholt10}, though non-detection of this alignment were also reported~\citep{tucker88,ulmer89}. For the satellite alignment, tidal torque is thought to play a major role in its formation~\citep{ciotti94,ciotti98,kuhlen07,pereira08,faltenbacher08,pereira10}. Its detection based on SDSS data has been reported by~\citet{pereira05, agustsson06,faltenbacher07}, while non-detections of this alignment are also reported based on data from both 2dF Galaxy Redshift Survey (2dFGRS)~\citep{colless01,bernstein02} and SDSS~\citep{siverd09}. In Table~\ref{table:summary}, we summarize the previous work that reports the existence and non-existence of these two types of alignments based on real data. On the other hand, the intrinsic alignment of galaxies will contaminate gravitational lensing measurements and therefore needs to be carefully modeled in lensing analysis. Along these lines,~\citet{mandelbaum06,hirata07} reported correlations between intrinsic shear and the density field based on data from SDSS and 2SLAQ~\citep{croom09}. 

In general, the cluster galaxy alignment signals are weak, and their measurement requires high quality photometry and well measured galaxy PAs. Moreover, a galaxy cluster catalog with high purity, well-determined BCGs and satellite galaxies is important too. Since measuring the redshift evolution of the alignment is crucial for understanding its origin, the cluster catalog needs to be volume limited and maintain constant purity for a wide redshift range. Most of the existing alignment measurements (see Table~\ref{table:summary}) are based on galaxy clusters/groups selected from spectroscopic data. In SDSS data, due to the high cost of obtaining spectra for a large population of galaxies, the completeness of spectroscopic coverage is limited to $r$ band Petrosian magnitude $r_p \le$17.7, corresponding to a median redshift of 0.1~\citep{strauss02}. This greatly limits the ability to look at the redshift evolution of the alignments.

On the other hand, one can also measure the alignments by using photometrically selected clusters. The advantage of photometrically selected clusters lies in the large data sample as well as relatively deep redshift coverage, allowing a study on the redshift evolution of the alignments. However, there are clear disadvantages too. For example, the satellite galaxies are prone to contamination from the projected field galaxies, which will dilute the alignment signals. The level of this contamination may also vary as redshift changes, complicating the interpretation of the alignment evolution. 

In this paper, we show our measurements of the two types of alignments based on a volume limited and highly pure ($\ge$ 90\%) subsample of clusters from the GMBCG cluster catalog for SDSS DR7~\citep{haocat}. The large sample of clusters allows us to examine the dependence of the alignments on cluster richness and redshift with sufficient statistics. With this catalog, we detect a BCG alignment that depends on redshift and the absolute magnitude of BCG, but not on the cluster richness. We also observe that the satellite alignment depends on the apparent brightness of the BCG and the methods the position angles are measured. We can only see a statistically significant satellite alignment at low redshift when we use the isophotal fit PAs (see \S~3.3 for more details). Furthermore, we notice that the satellite alignment based on isophotal fit PAs depends strongly on the apparent magnitude rather than the absolute magnitude of the BCG. This suggests that the measured satellite alignment is more likely due to the isophotal fit PAs, whose measurements are prone to contaminations from the diffuse light of the BCG.  

The paper is organized as following: in \S 2, we introduce the two parameters used to quantify the two types of alignments. In \S~3, we introduce the data used in this paper. In \S~4, we present and discuss our measurement results. By convention, we use a $\Lambda$CDM cosmology with $h=1.0$, $\Omega_m=0.3$ and $\Omega_{\Lambda}=0.7$ throughout this paper. All angles measurements that appear in this paper are in units of degrees. 

\begin{deluxetable*}{l l|l|l|r}

\tabletypesize{\tiny}
\tablecolumns{5} 
\tablewidth{0pt} 
\tablecaption{Alignment Measurements Summary\label{table:summary}}
\tablehead{
	   \multicolumn{1}{c}{Data }&
	   \multicolumn{2}{c}{BCG Alignment}&
           \multicolumn{2}{c}{Satellite Alignment}\\ 
           \colhead{Source}& \colhead{Exist}& \colhead{Non-exist} & \colhead{Exist} & \colhead{Non-exist}	
}
\startdata 
               &\citet{sastry68}   &\citet{tucker88}   &                        &\\
               &\citet{austin74}   &\citet{ulmer89}    &                        &\\
  Photometric  &\citet{dressler78} &                   &                        &\\
  Plates       &\citet{carter80}   &                   &                        &\\
               &\citet{binggeli82} &                   &                        &\\ \tableline
               &\citet{brainerd05} &                   &\citet{pereira05}       & \citet{siverd09}\\
 SDSS          &\citet{yang06}     &                   &\citet{agustsson06}     &\\
 Spectrosopic  &\citet{azzaro07}   &                   &\citet{faltenbacher07}  &\\ 
               &\citet{faltenbacher07} &               &                        &\\ 
	       &\citet{wangyougang08}  &               &                        &\\ 
               &\citet{siverd09}   &                   &                        &\\ \tableline
 SDSS          &\citet{ostholt10}  &                   & \citet{pereira05}      &\\ 
 Photometric   &\color{red}{This Work}          &      &                        &\color{red}{This Work}\\ \tableline
 2dF           &                   &                   &                        &\citet{bernstein02}\\
 Spectroscopic &                   &                   &                        &           
\enddata
\end{deluxetable*}

\section{Alignment parameters}
\begin{figure*}
\epsscale{0.9}
\begin{center}
\plottwo{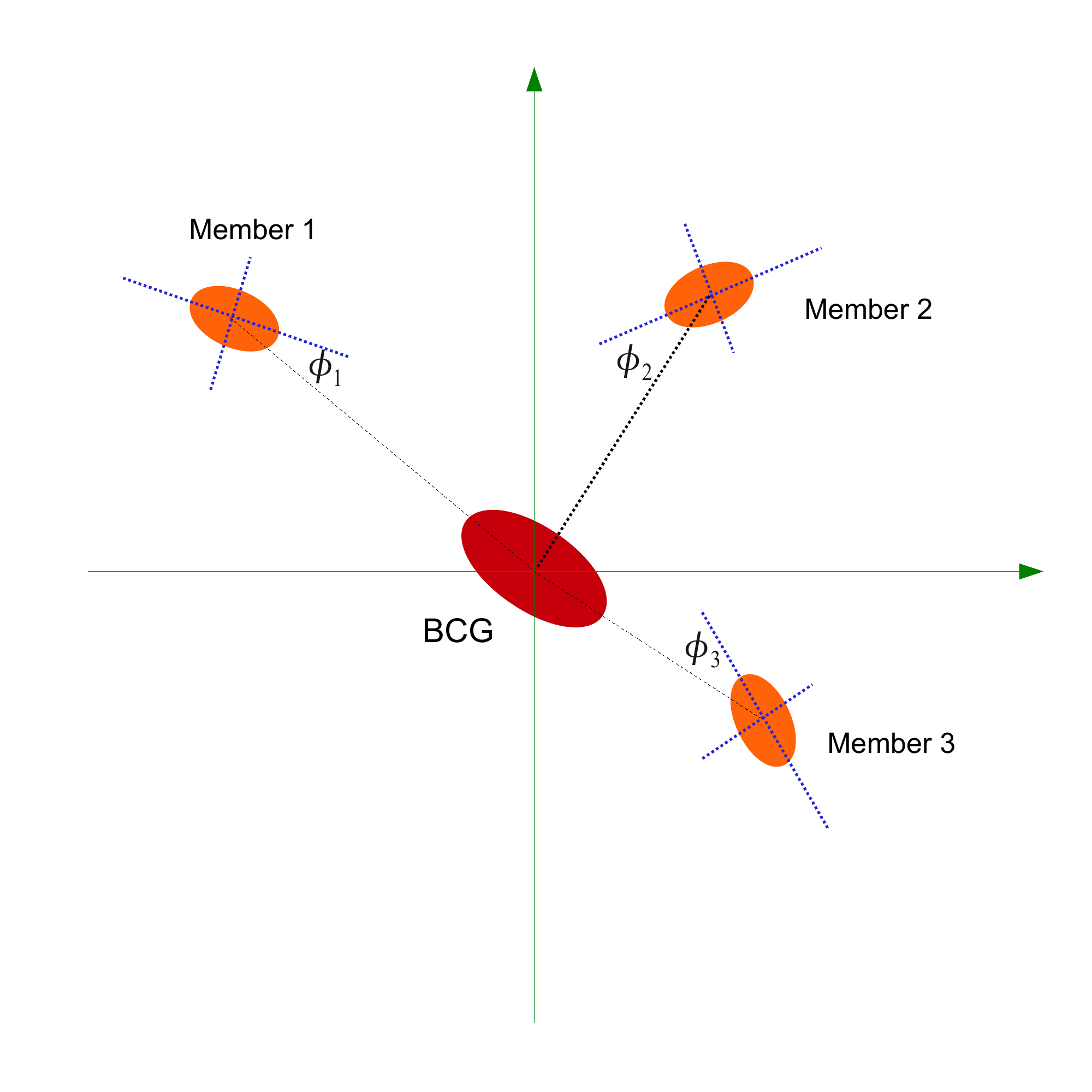}{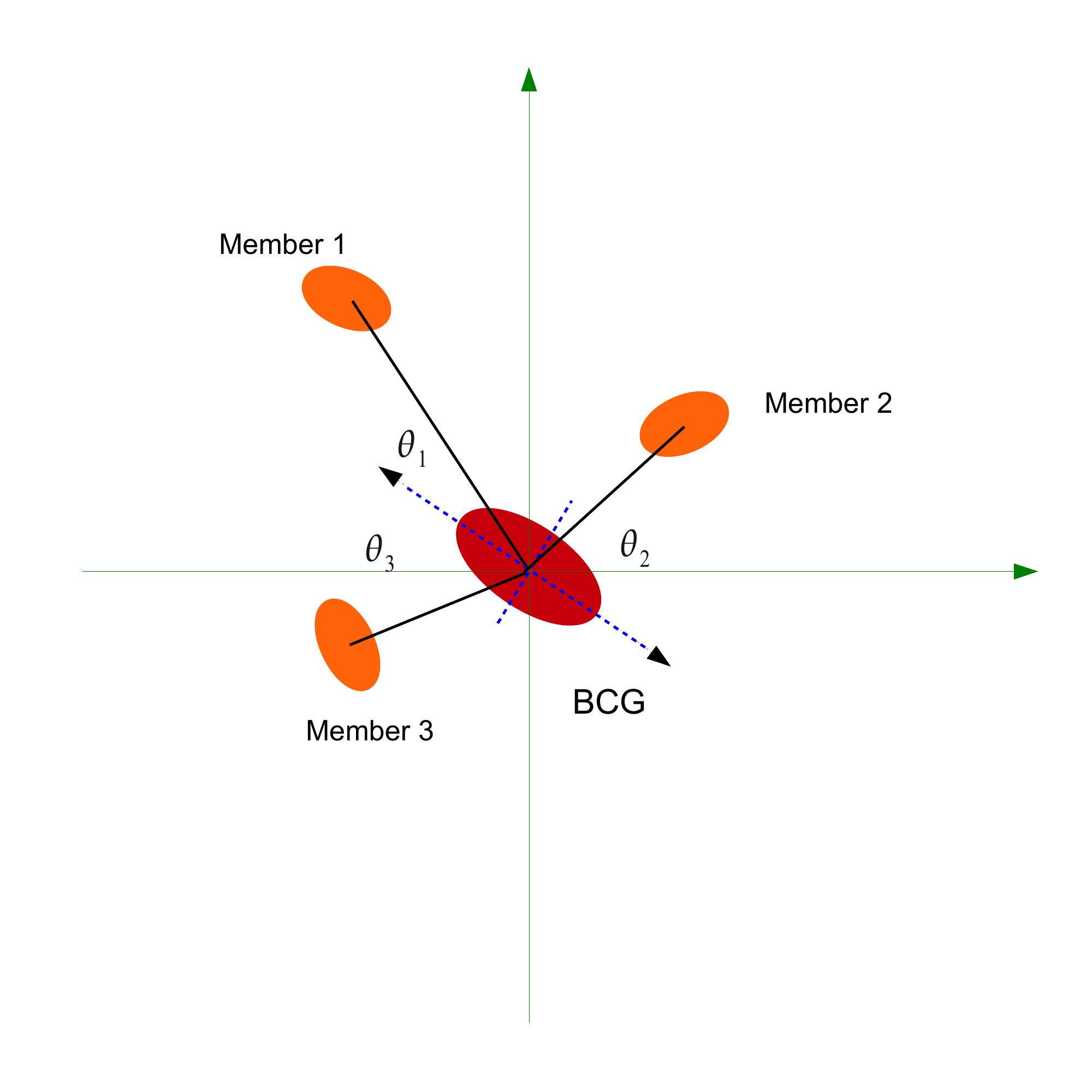}
\caption{\textit{left:} Illustration of the angle $\phi$ used in the definition of satellite alignment parameter $\delta$; \textit{right:} Illustration of the angle $\theta$ used in the definition of BCG alignment parameter $\gamma$. }
\label{fig:aangle}
\end{center}
\end{figure*}

In this paper, we consider two types of alignments: (1) Satellite alignment; (2) BCG alignment. Each of them is quantified by a corresponding alignment parameter. For the satellite alignment, we follow~\citet{struble,pereira05} and use the following alignment parameter:

\begin{equation}\label{eq:delta}
\delta = \frac{\sum^{N}_{i=1}{\phi_i}}{N} - 45
\end{equation}

\noindent where $\phi_i$ is the angle between the major axes of the satellite galaxies and the lines connecting their centers to the BCGs, as illustrated in the left panel of Figure~\ref{fig:aangle}. N is the number of satellite galaxies in the cluster. For every cluster, there will be a unique alignment parameter $\delta$ measured, which is the mean angle $\phi$ of all cluster satellite galaxies subtracted by 45. If the major axes of satellite galaxies do not preferentially point to the BCG of the cluster, the $\phi$ will randomly distribute between -45 and 45 (degrees), leading to $\delta = 0$. On the other hand, if the major axes of satellite galaxies preferentially point to the BCG, there will be $\delta < 0$. The standard deviation (error bar) of $\delta$ can be naturally calculated by $\delta_{err} = \sqrt{\sum_{i=1}^N{(\phi_i-\delta-45)^2}}/N$.

For the BCG alignment, we will focus on the angle $\theta$ between the BCGs PA and the lines connecting the BCG to each satellite galaxy, as illustrated in the right panel of Figure~\ref{fig:aangle}. In our cluster sample, each cluster has more than 15 satellite galaxies (see \S 3.2 for more details). Therefore, instead of looking at the full distribution of $\theta$ from all clusters, we will focus on the mean of $\theta$ measured for each cluster. In analogy to the satellite alignment parameter $\delta$, we introduce a BCG alignment parameter $\gamma$ defined as:

\begin{equation}\label{eq:gamma}
\gamma = \frac{\sum^{N}_{i=1}{\theta_i}}{N} - 45
\end{equation}

\noindent where $\theta_i$ is the angle between BCGs PA and the line connecting the BCG to the $i^{th}$ satellite galaxy (see the right panel of Figure~\ref{fig:aangle}). N is the number of satellite galaxies in the cluster. Similarly, each cluster will correspond to a BCG alignment $\gamma$, which is the mean of angle $\theta$ subtracted by 45. If the major axis of BCG preferentially aligns with the distribution of the majority of the satellite galaxies, the $\gamma < 0$. If no such preference exist, $\gamma = 0$ will be expected. The uncertainty of $\gamma$ can be readily calculated as $\gamma_{err} = \sqrt{\sum_{i=1}^N{(\theta_i-\gamma-45)^2}}/N$.

These two parameters quantify the two types of alignments and are easy to measure. In the follows, we will focus on these two quantities and their dependencies on various cluster/BCG properties.

\section{Data}
In this section, we describe the details of the galaxies, their PA measurements, and the galaxy cluster sample used in our measurement.

\subsection{Galaxy Catalog}
 
The galaxies we use are from the Data Release 7 of the Sloan Digital Sky Survey~\citep{york00, abazajian08}. The SDSS is a multi-color digital CCD imaging and spectroscopic sky survey, utilizing a dedicated 2.5-meter telescope at Apache Point Observatory, New Mexico. It has recently completed mapping over one quarter of the sky in $u$,$g$,$r$,$i$ and $z$ filters. DR7 is a mark of the completion of the original goals of the SDSS and the end of the phase known as SDSS-II. It includes a total imaging area of 11663 square degrees with 357 million unique objects identified.

This work focuses on the Legacy Survey area of SDSS DR7, which covers more than 7,500 square degrees of the North Galactic Cap, and three stripes in the South Galactic Cap totaling 740 square degrees~\citep{abazajian08}. The galaxies are selected from the PhotoPrimary view of the SDSS Catalog Archive Server with object type tag set to 3 (galaxy) and $i$-band magnitude less than 21.0. Moreover, we require that the galaxies did not trigger the following error flags: SATURATED, SATUR\_CENTER, BRIGHT, AMOMENT\_MAXITER, AMOMENT\_SHIFT and AMOMENT\_FAINT. 

In addition to the above selection criteria, we also reject those galaxies with photometric errors in $r$ and $i$ band greater than 10 percent. Additionally, we require the ellipticity\footnote{The ellipticity is defined as $\mathrm{\sqrt{m_{e1}^2 +m_{e2}^2}}$, where the $\mathrm{m_{e1} = \frac{<col^2> - <row^2>}{<col^2> + <row^2>}}$ and $\mathrm{m_{e2} = \frac{2 <col*row>}{<col^2> + <row^2>}}$. Details of estimating $\mathrm{<...>}$ can be found in \citep{bernsteinjarvis02}} of each galaxy in the $r$-band and $i$-band to be less than 0.8 in order to remove edge-on galaxies whose colors are not well measured. By doing this, we will retain about 95\% of the total galaxies. All the magnitudes used in this paper are dust extinction corrected model magnitudes~\citep{abazajian08}. 

\subsection{Galaxy Clusters}
\begin{figure*}
\begin{center}
\epsscale{0.8}
\plotone{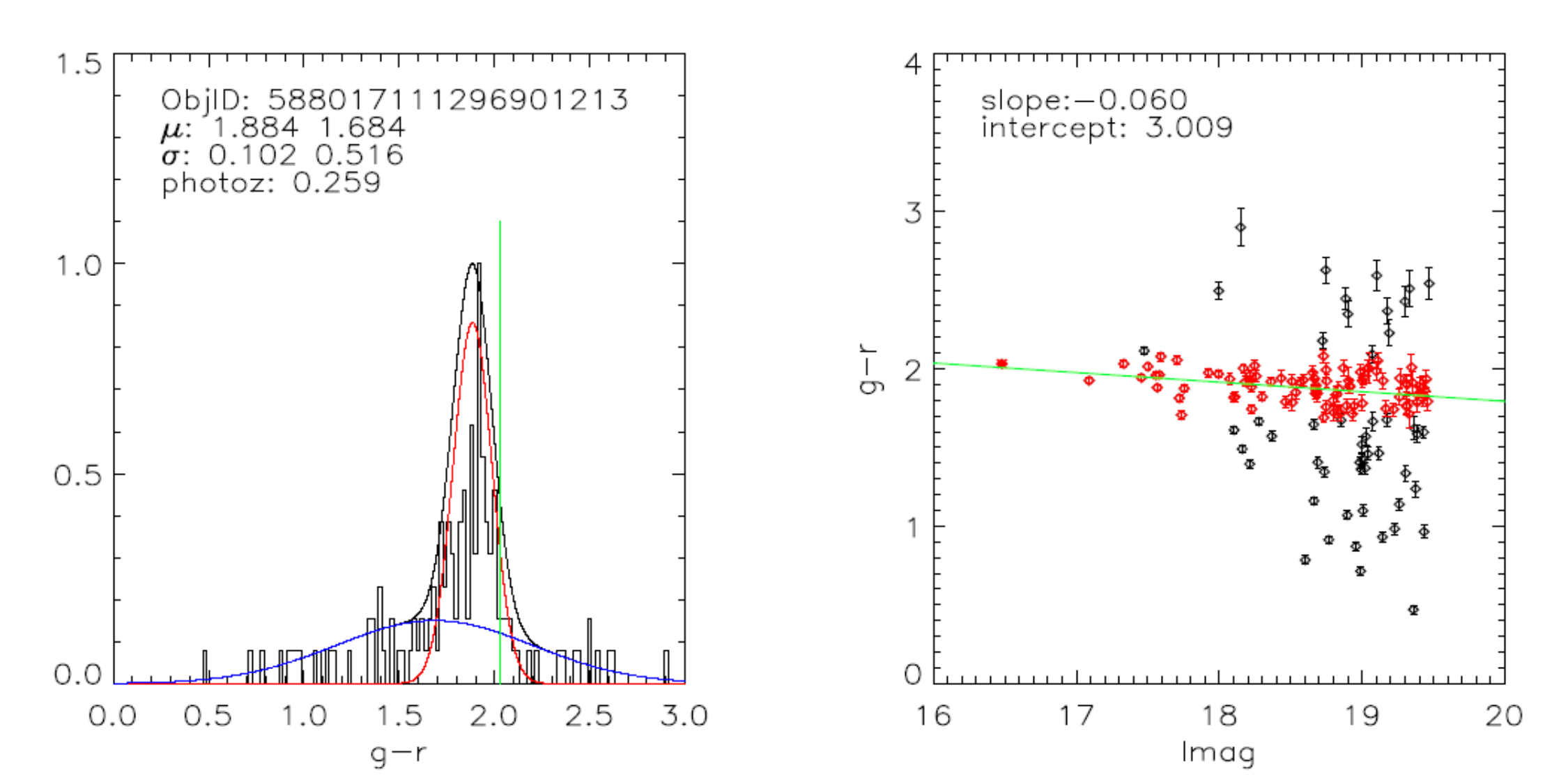}
\caption{{\em left:} Galaxy $g-r$ color distribution around a cluster
  overlaid with a model constructed of a mixture of two Gaussian
  distributions. The red curve corresponds to the red sequence
  component while the blue one corresponds to the sum of background
  galaxies and blue cluster satellites. The green vertical line indicates
  the color of the BCG. $\mu$ and $\sigma$ are the means and standard
  deviations of the two Gaussian components. {\em right:}
  Color-magnitude relation for the same galaxies.  Galaxies within the
  2$\sigma$ clip of the red sequence component are shown with red
  points; the green line indicates the best fit slope and intercept of
  this red seqence.}
\label{fig:gmr_bimodal}
\end{center}
\end{figure*}

In order to measure the two types of alignments, we need to have a galaxy cluster catalog with well determined member galaxies. The GMBCG cluster catalog for SDSS DR7 is a large catalog of optically selected clusters from SDSS DR7 using the GMBCG algorithm. The catalog is constructed by detecting the BCG plus red sequence feature that exists among most clusters. The cluster satellite galaxies are within 2$\sigma$ of the red sequence mean color detected using a Gaussian Mixture Model. Since the red sequence of each cluster is measured individually, it allows a more accurate satellite galaxy selection than using a universal red sequence model. We count the satellite galaxies down to 0.4L* at the cluster's redshift. In Figure~\ref{fig:gmr_bimodal}, we show the color distribution of galaxies around a cluster in GMBCG catalog. To study the redshift evolution of the alignment, we choose the volume limited sample of clusters with redshift below 0.4, where the satellite galaxies are selected using $g-r$ color. Also, to get high purity, we choose clusters with richness equal to or greater than 15, which leads to a purity above 90\% across the redshift range. As a result, there are about 11,000 clusters with over 260,000 associated satellite galaxies. More details of the cluster catalog can be found in~\citet{haocat}. 

\subsection{PA Measurements}
In the SDSS data reduction pipeline, the PAs of galaxies are measured with several different methods~\citep{stoughton02}. In this work, we will use three of them: isophotal PA, exponential fit PA~\footnote{Since most satellite galaixes are selected using red sequence, the exponential fit may not be suitable. In this paper, we want to demonstrate how will the PA measurement affect the alignment signal, and therefore include it in our discussion.} and De Vaucouleurs fit PA. 

To measure the isophotal PA, the SDSS pipeline measures out to the 25 magnitudes per square arc-second isophote (in all bands). The radius of a particular isophote as a function of angle is measured and Fourier expanded. From the coefficients, PA (isoPhi) together with the centroid (isoRowC,isoColC), the major and minor axes (isoA,isoB) are extracted.

\begin{figure*}
\epsscale{0.75}
\begin{center}
\plottwo{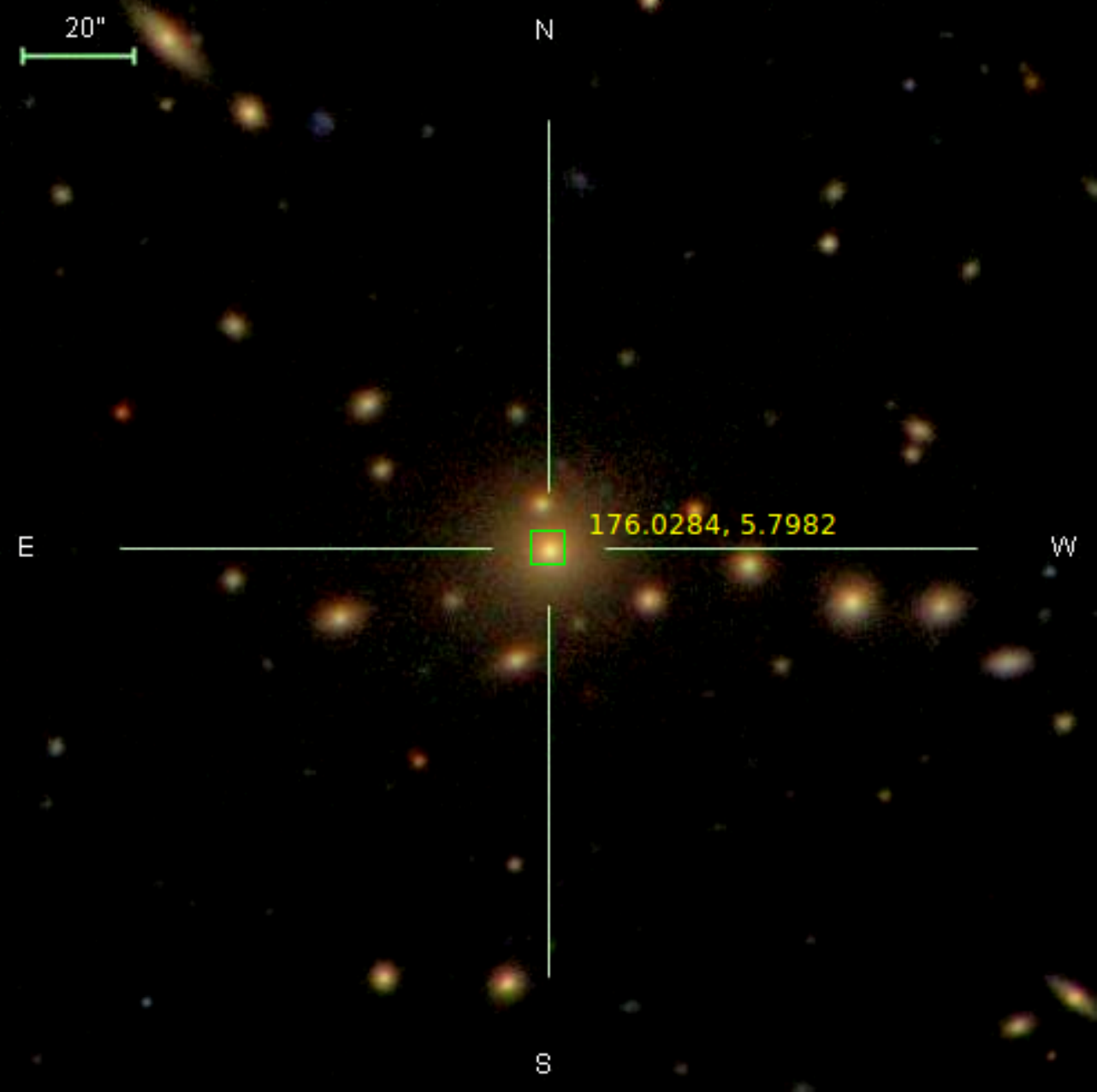}{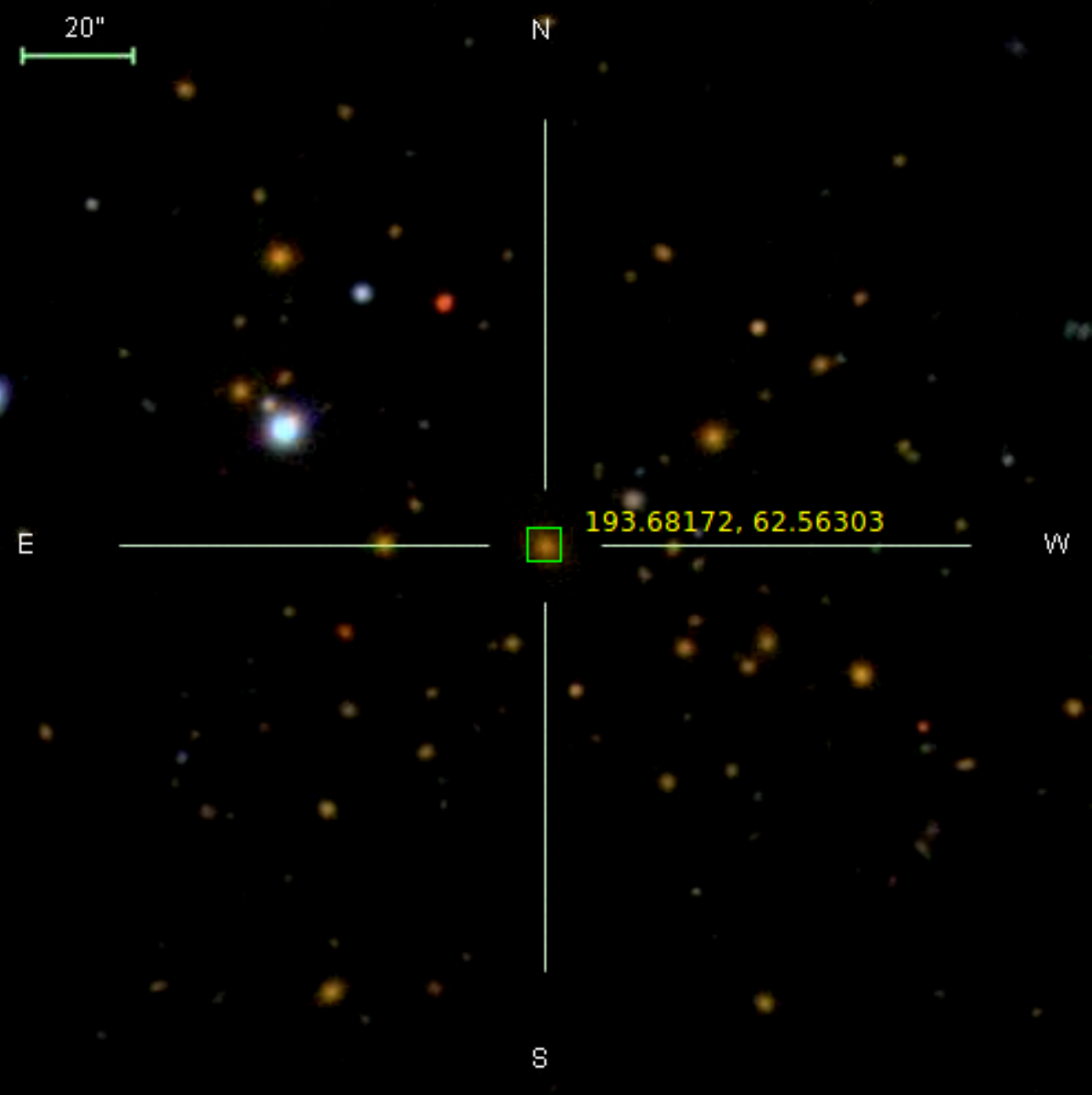}
\caption{Left is a cluster at redshift 0.103 and the right is a cluster at redshift 0.310. The light from BCG will pose a risk on the proper measurements of the PAs of the satellite galaxies, especially at low redshift.}
\label{fig:diffuse}
\end{center}
\end{figure*}

For the exponential fit PA, the SDSS pipeline fit the intensity of galaxy by 
\begin{equation}
I(R)=I_0\exp[-1.68 (R/R_{eff})]
\end{equation}
\noindent where the profile is truncated outside of 3$R_{eff}$ and smoothly decreases to zero at 4$R_{eff}$. After correcting for the PSF, the PA is calculated from this fitting and reported as expPhi in the CASJOB database. 
The De Vaucouleurs fit PA follows the same procedure as exponential fit PA, except the model in the fitting is 
\begin{equation}
I(R)=I_0\exp[-7.67 (R/R_{eff})^{1/4}]
\end{equation}

\noindent where the profile is truncated outside of 7$R_{eff}$ and smoothly decreases to zero at 8$R_{eff}$. The PA from this fitting is reported as devPhi. For more details about these PA measurements, one can refer to~\citet{stoughton02}. All the angles are in degrees East of North by convention. By comparing these different PAs, isophotal PA tends to trace the exterior shape of the galaxy while the two model fit PAs tend to trace the inner profile of the galaxy. At low redshift, the BCG is very bright and its diffuse light may severely affect the measurement of the outer part of the nearby galaxies. Therefore, the isophotal PA is more susceptible to this artifact, leading to an ``artificial'' orientation preference toward BCG. To give a sense of the BCG diffuse light effect, we show a low redshift and a higher redshift cluster in Figure~\ref{fig:diffuse}.

\subsection{Control Sample}

As we talk about the detection of the alignment signals, we need to have a control sample to compare with. Naively, one may directly compare the measured alignment signal to what is expected for non-detection. However, this will implicitly assume that the data as well as the measurements are free from any systematics. When talking about the significance level of the detection, it is not sufficient to consider only the statistical uncertainties. Systematics (intrinsic scatter), if they exist, also need to be considered. Therefore, introducing an appropriate control sample and applying the same measurements we apply to the cluster data will help to eliminate the possible false detection resulted from potential systematics. 

For this purpose, we prepare our control sample as follows: we shuffle the BCGs by assigning random positions (RA and DEC) to them, but keep all other information of the BCGs unchanged. Then around each BCG (at new random position), we re-assign ``cluster satellites'' by choosing those galaxies that are falling within the $R_{scale}$ from the BCG. Also, the $i$ band magnitude of the ``satellites'' should be in the range from 14 to 20. The $R_{scale}$, measured in Mpc, plays the role of virial radius~\citep{haocat}. In addition to this non-cluster sample, we also add another random PA to every galaxy in both the true cluster sample and the non-cluster sample by replacing the measured galaxy's PA with a random angle uniformly sampled between 0 and 180 degree. This random PA will serve as another control sample to double check the possible systematics in our measurements. 

\section{Results}

\subsection{Satellite Alignment}
\begin{figure}
\begin{center}
\epsscale{1}
\plotone{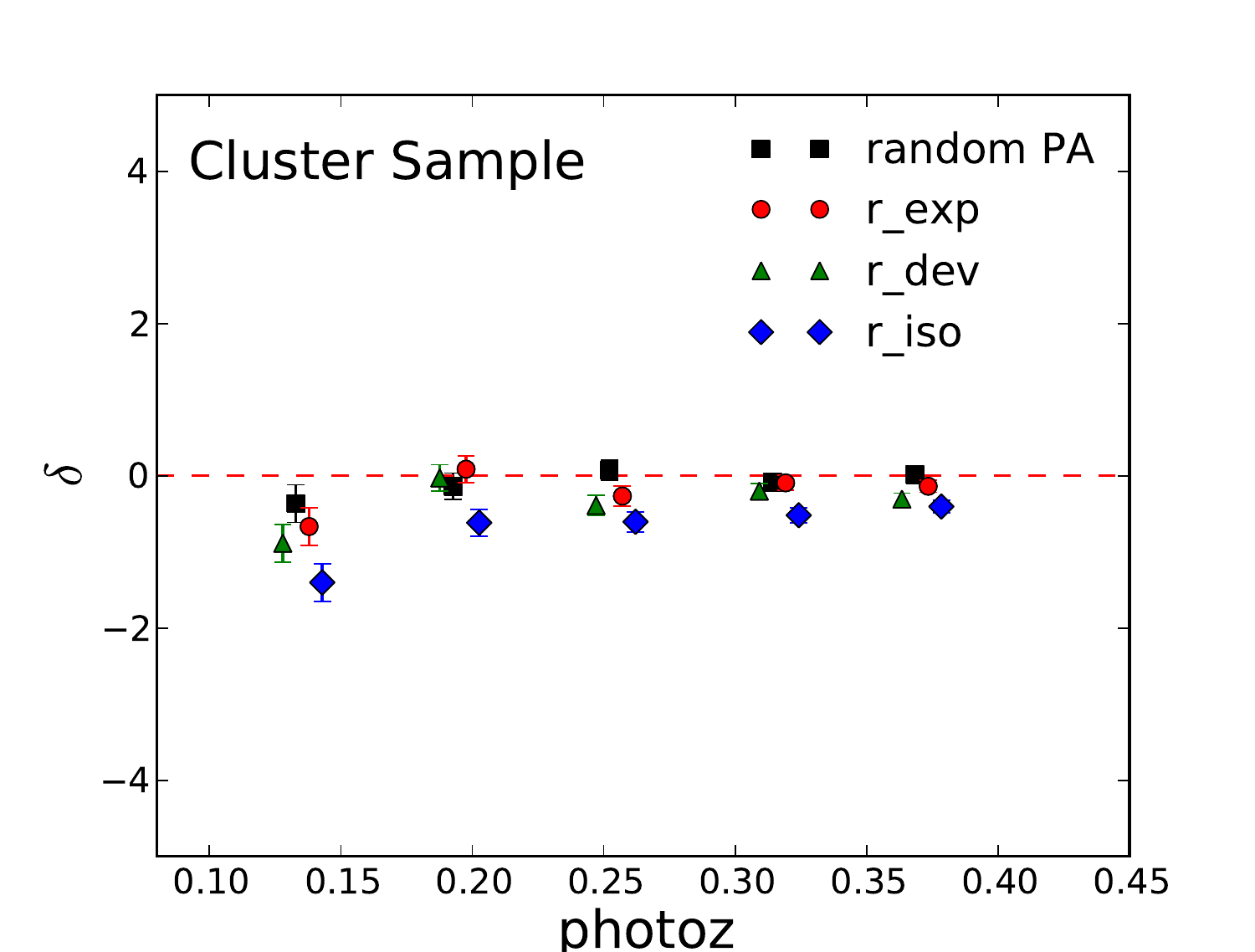}
\caption{Satellite alignment for clusters at different redshift bin of size 0.05. The legend random PA indicates using of randomized PAs. exp indicates using exponential fit PAs. dev indicates using De Vaucouleurs fit PAs and iso indicates using isophotal PAs. $r$ indicates the SDSS $r$ filter. This legend convention is also applicable to other figures in the paper.}
\label{fig:deltaz}
\end{center}
\end{figure}

\begin{figure}
\begin{center}
\epsscale{1}
\plotone{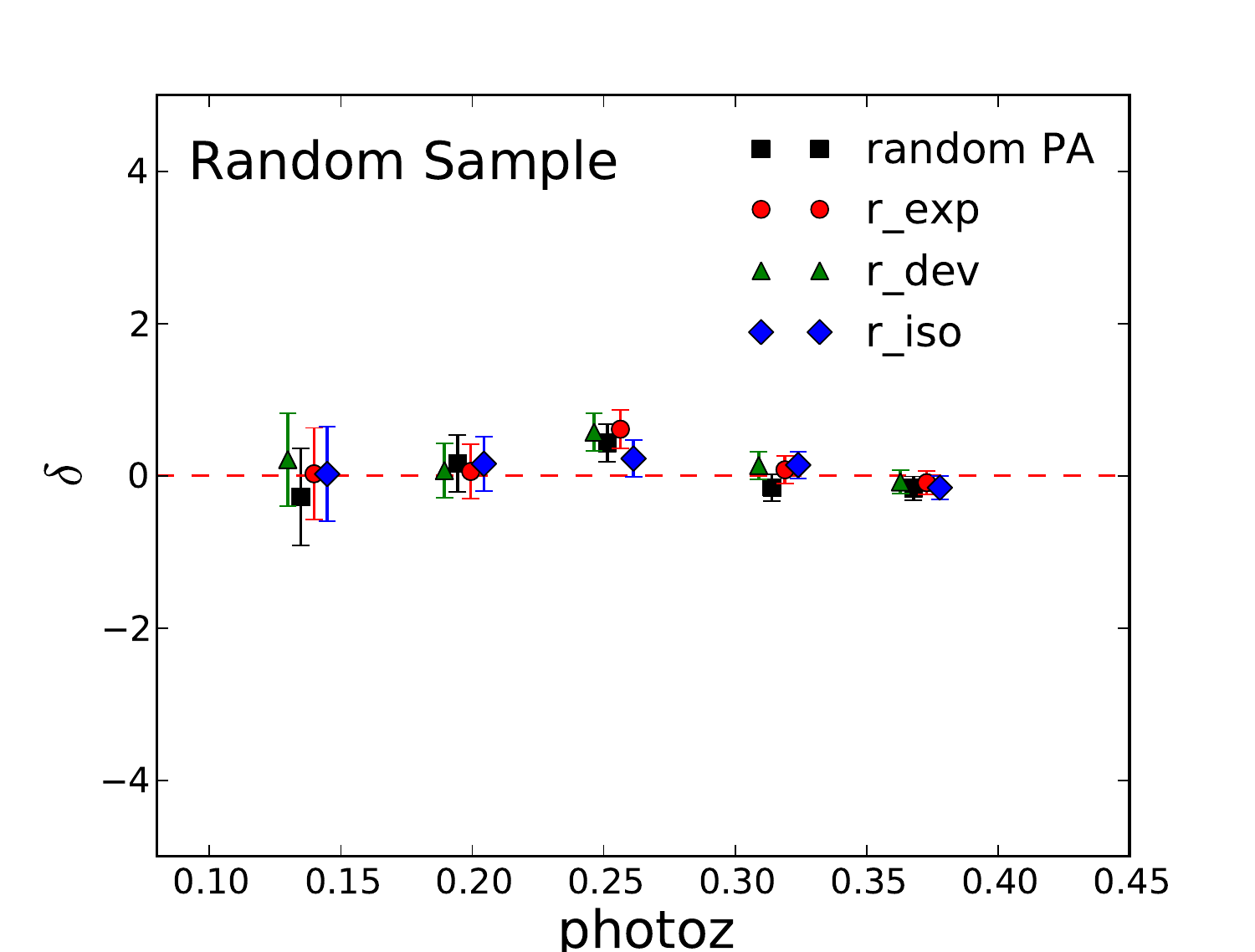}
\caption{Satellite alignment measured based on the random control sample. }
\label{fig:delta_z_random}
\end{center}
\end{figure}

There are two basic questions concerning the formation of the satellite alignment: (1) is it a residual feature of the initial condition of the cluster formation? (2) or is it a dynamically evolving effect that varies as the cluster evolves~\citep{pereira05}, for example, due to the tidal torque? The two different scenarios lead to different redshift dependence of $\delta$. If it is left over from the initial alignment, its strength should decrease as redshift decreases. On the other hand, we should see a stronger alignment signal at low redshift if it is a dynamically evolving effect~\citep{ciotti94,catelan00,kuhlen07,pereira08,pereira10}. Therefore, looking at the redshift dependence of the measured $\delta$ is our primary interest. To do this, we first measure the $\delta$ for each cluster using 4 different PAs, i.e., the random PA, exponential fit PA, De Vaucouleurs fit PA and isophotal PA in $r$ band. We bin the clusters into redshift bins of size 0.05. In each bin, we calculate the weighted mean of $\delta$ and the standard deviation of the weighted mean, with the weights specified by $1/\delta_{err}^2$. Then, we perform the same measurement on the control sample, and present the results in Figure~\ref{fig:deltaz} and Figure~\ref{fig:delta_z_random}.

\begin{figure}
\begin{center}
\epsscale{1}
\plotone{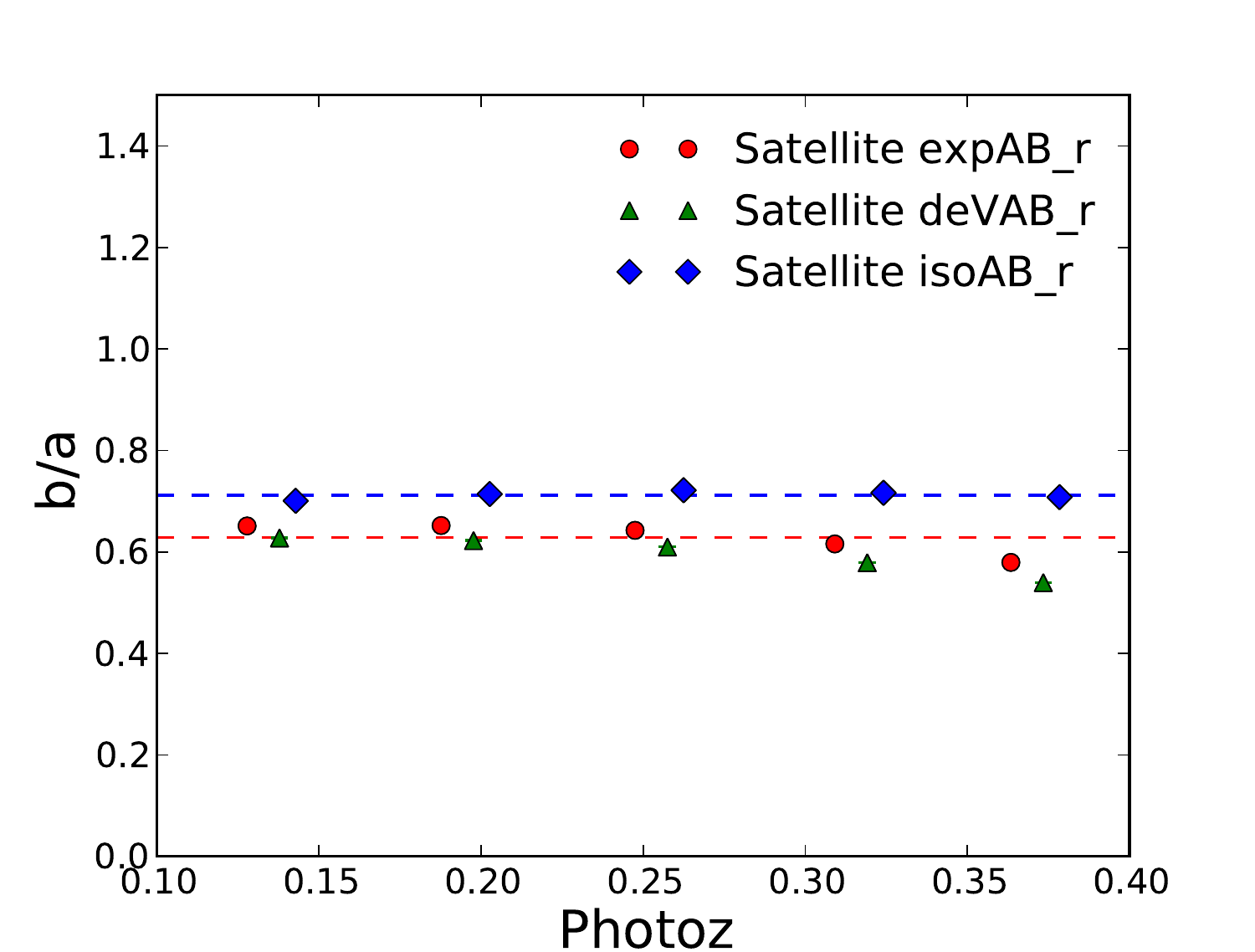}
\caption{The axis ratio of the satellite galaxies. Here, the error bar is the standard deviation to the mean in that bin. The legend satellite expAB\_r refers to the $b/a$ ratio is measured by fitting the satellite galaxy with an exponential profile in $r$ band. devAB\_r and isoAB\_r refer to the $b/a$ ratio by fitting De Vaucouleurs and isophotal profiles respectively. This convention is also applicable to other figures in this paper.}
\label{fig:deltambz}
\end{center}
\end{figure}

Based on the results in Figure~\ref{fig:deltaz}, one can see that the $\delta$ measured using isophotal PAs deviated from that based on random PAs in a statistically significant way at low redshift, though approaching zero as redshift increases. While the $\delta$ measured using the exponential fit PA and De Vaucouleurs fit PA are consistent with that measured using random PAs except in the lowest redshift bin. In \citet{pereira05}, the authors used isophotal fit PAs and also considered a cluster sample (sample B) with satellite galaxies selected using red sequence. When limiting the $r$ band magnitude to less than 18, they measured a $\delta = -1.06 \pm 0.37$, which is consistent with our results in the lowest redshift bin using isophotal fit PAs. The results based on the random control sample in Figure~\ref{fig:delta_z_random} show that $\delta$ measured using all types of PAs are consistently zero across the redshift range. In Figure~\ref{fig:deltambz}, we plot the axis ratio of the satellite galaxies in different redshift bins. The axis ratio determines how precisely the PAs are measured. From Figure~\ref{fig:deltambz}, the axis ratio measured using the isophotal method does not vary much as redshift increases, indicating that the diminishing satellite alignment based on isophotal fit PAs is not due to the decreasing S/N at high redshift. 

There are two possible explanations for the measured $\delta$ in the low redshift bins. The first one is that the diffuse light from the BCGs affects the measurements of the PAs of the cluster satellite galaxies'. This creates an artificial preference of major axes of the satellite galaxies. This contamination is most severe when the PAs are measured using isophotal fit, but less prominent when the PAs are measured using exponential fit and De Vaucouleurs fit. This is because the isophotal PAs are sensitive to the shape of the outer profile of galaxy while the model fit PAs are more determined by the inner profile of the galaxy. 

\begin{figure}
\epsscale{1}
\plotone{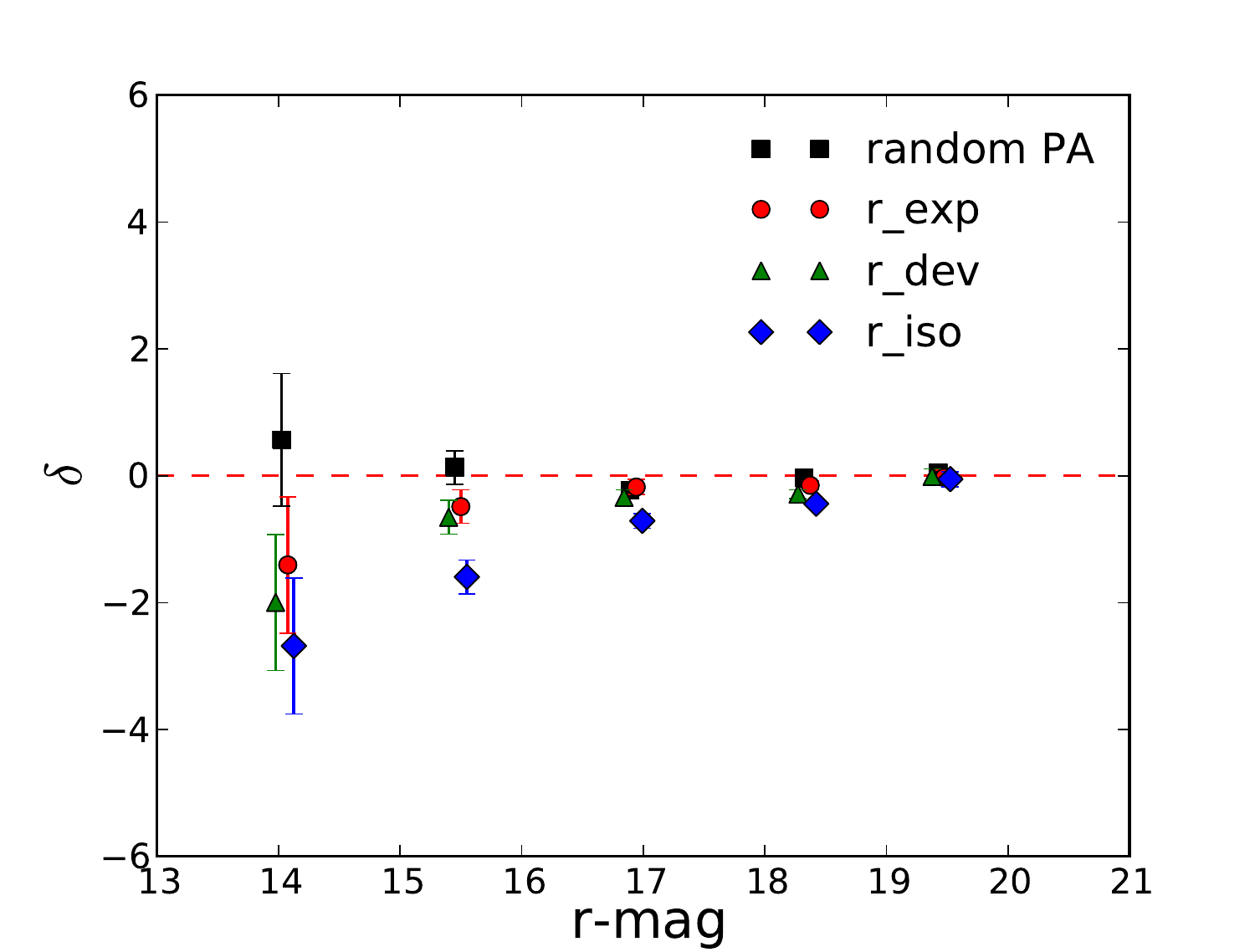}
\caption{The dependence of satellite alignment $\delta$ on the r-band apparent magnitudes of the corresponding BCGs. This shows a strong dependence of $\delta$ measured using isophotal PAs on the apparent magnitude of the BCGs.}
\label{fig:deltarmaga}
\end{figure}

\begin{figure}
\epsscale{1}
\plotone{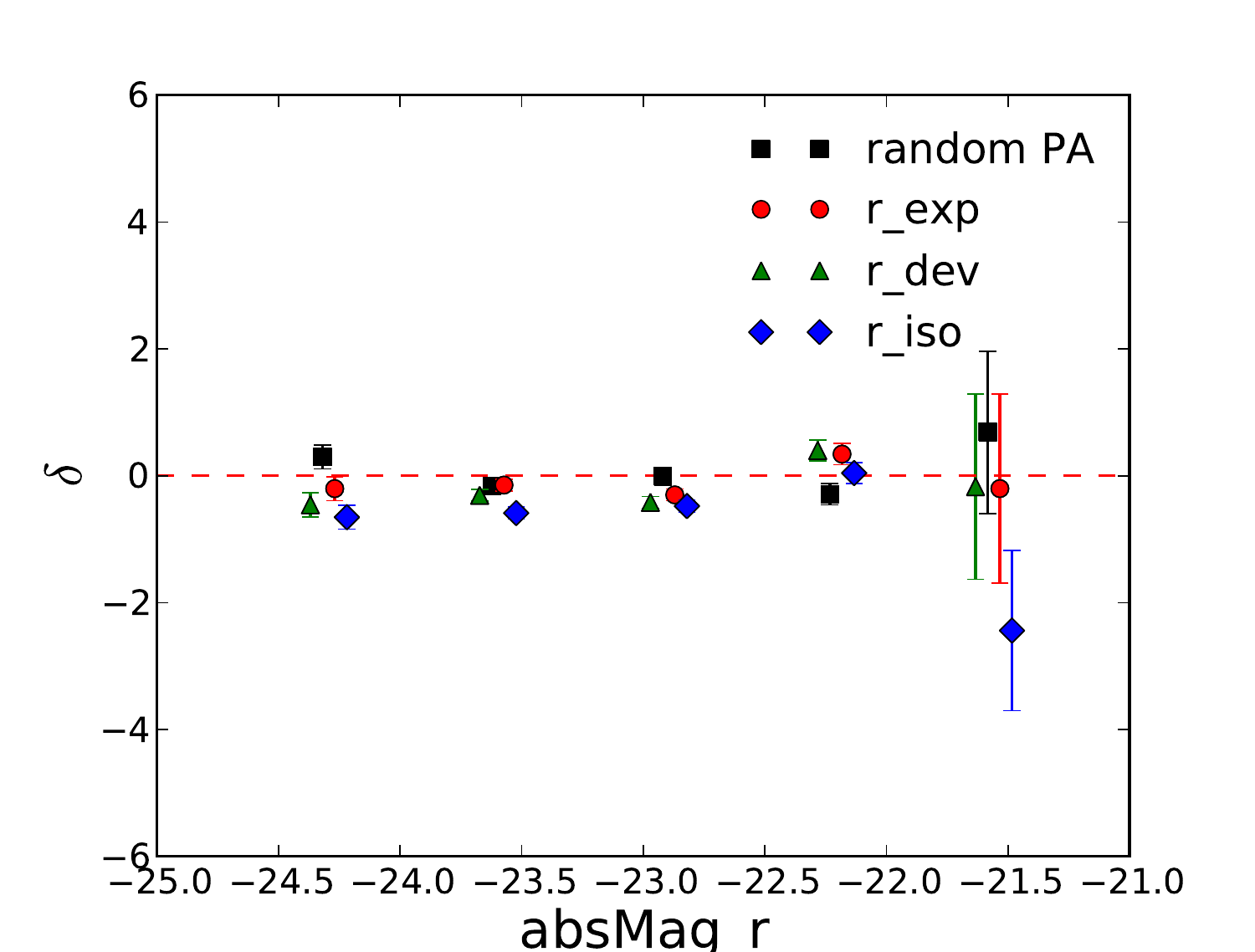}
\caption{The dependence of satellite alignment $\delta$ on the r-band absolute magnitude of the BCGs. This shows that there is very little dependence of $\delta$ measured using isophotal PAs on the $r$ band absolute magnitude of the BCGs}
\label{fig:deltarmagb}
\end{figure}

\begin{figure}
\epsscale{1}
\plotone{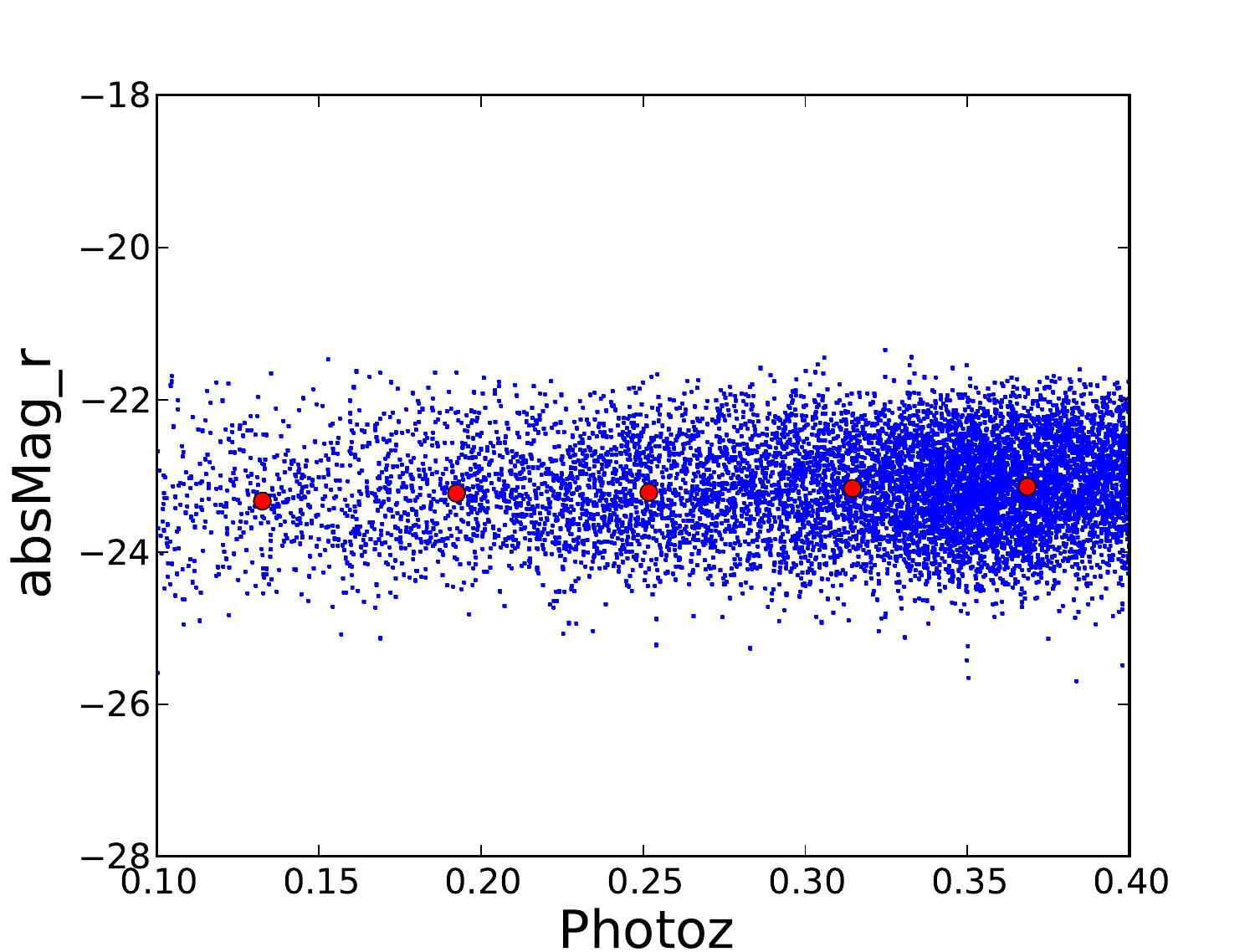}
\caption{Photometric redshift vs. r-band absolute magnitude for BCGs. The red overplotted dots are the means in each redshift bin of size 0.05.}
\label{fig:photozramag}
\end{figure}

The second possible explanation to these results is the twisting of galaxy. This leads to different PAs when we use different methods. The outer rim of the galaxy is more susceptible to the tidal torque so that the alignment will show up when we use the isophotal fit PAs. One way to distinguish these two explanations is to look at the way $\delta$ depends on the absolute and apparent magnitudes of the BCG. To see this, we plot the measured $\delta$ for all the clusters with respect to apparent magnitudes and absolute magnitudes of the corresponding BCGs in the Figure~\ref{fig:deltarmaga} and Figure~\ref{fig:deltarmagb} respectively. One can see that $\delta$ shows a strong dependence on the apparent magnitude but not on the absolute magnitude. Therefore, we conclude that the $\delta$ in the low redshift bins in Figure~\ref{fig:deltaz} is more likely resulted from the artifact of the PA measurement. In Figure~\ref{fig:photozramag}, we show the absolute magnitude vs redsfhit for the BCGs. 

\subsection{BCG Alignment}

\subsubsection{Redshift Dependence}
\begin{figure}
\begin{center}
\epsscale{1}
\plotone{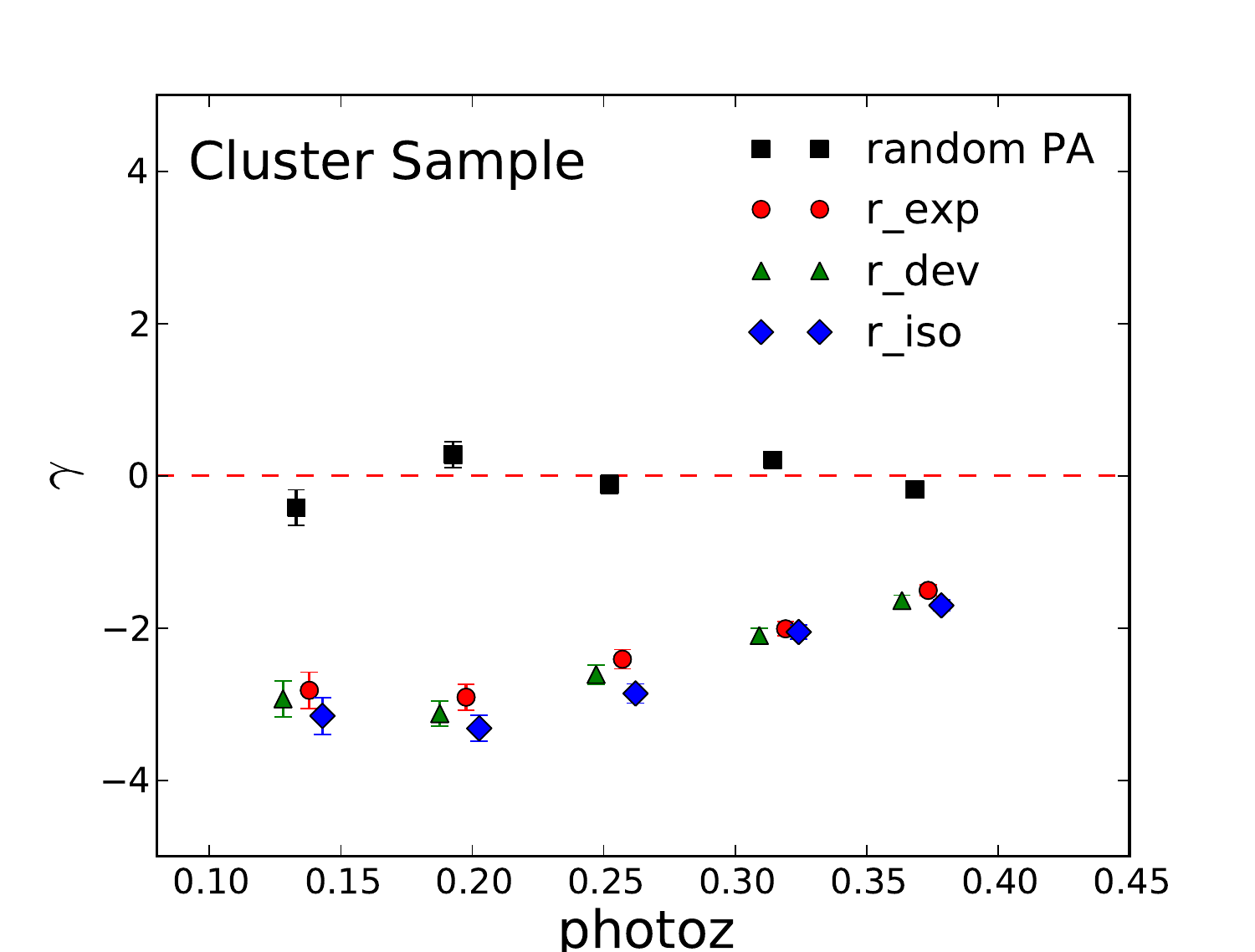}
\caption{BCG alignment at redshift bins of size 0.05 using different PAs measured from the cluster sample.}
\label{fig:gammaz}
\end{center}
\end{figure}

\begin{figure}
\begin{center}
\epsscale{1}
\plotone{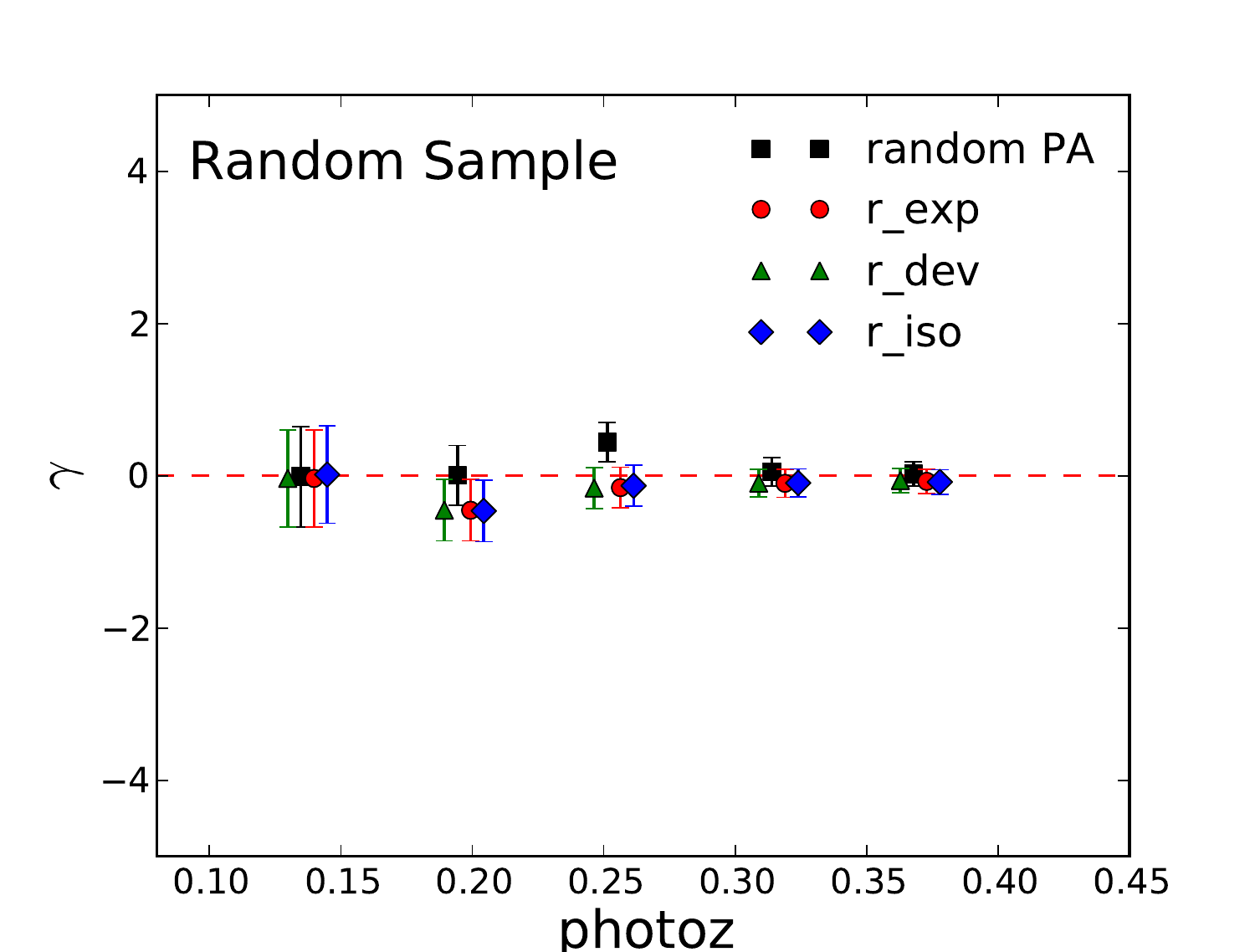}
\caption{BCG alignment at redshift bins of size 0.05 using different PAs measured from the random sample.}
\label{fig:gammaz_random}
\end{center}
\end{figure}

Now, we consider the BCG alignment. We first look at the redshift dependence of $\gamma$. Again, we perform our measurements on both the cluster sample and the random control sample. The results are shown in Figure~\ref{fig:gammaz} and Figure~\ref{fig:gammaz_random}

The results from the random control sample show that $\gamma$ is consistently zero. From the cluster sample, we detect a clear BCG alignment and its strength decreases as redshift increases. From the figures, we can also see that $\gamma$ is almost the same no matter how the PAs of BCGs  are measured. This has two important implications: 1. the PAs of BCGs are well measured by both isophotal fit and model fit; 2. the diffuse light of satellite galaxies does not affect the PA measurement of the BCG. Before we can conclude the redshift dependence of $\gamma$, we still need to do one more test. That is, if the axis ratio ($b/a$) of the BCG and cluster become systematically smaller due to the decreased S/N at higher redshift, the measured strength of $\gamma$ will decrease too. 

\begin{figure}
\epsscale{1}
\plotone{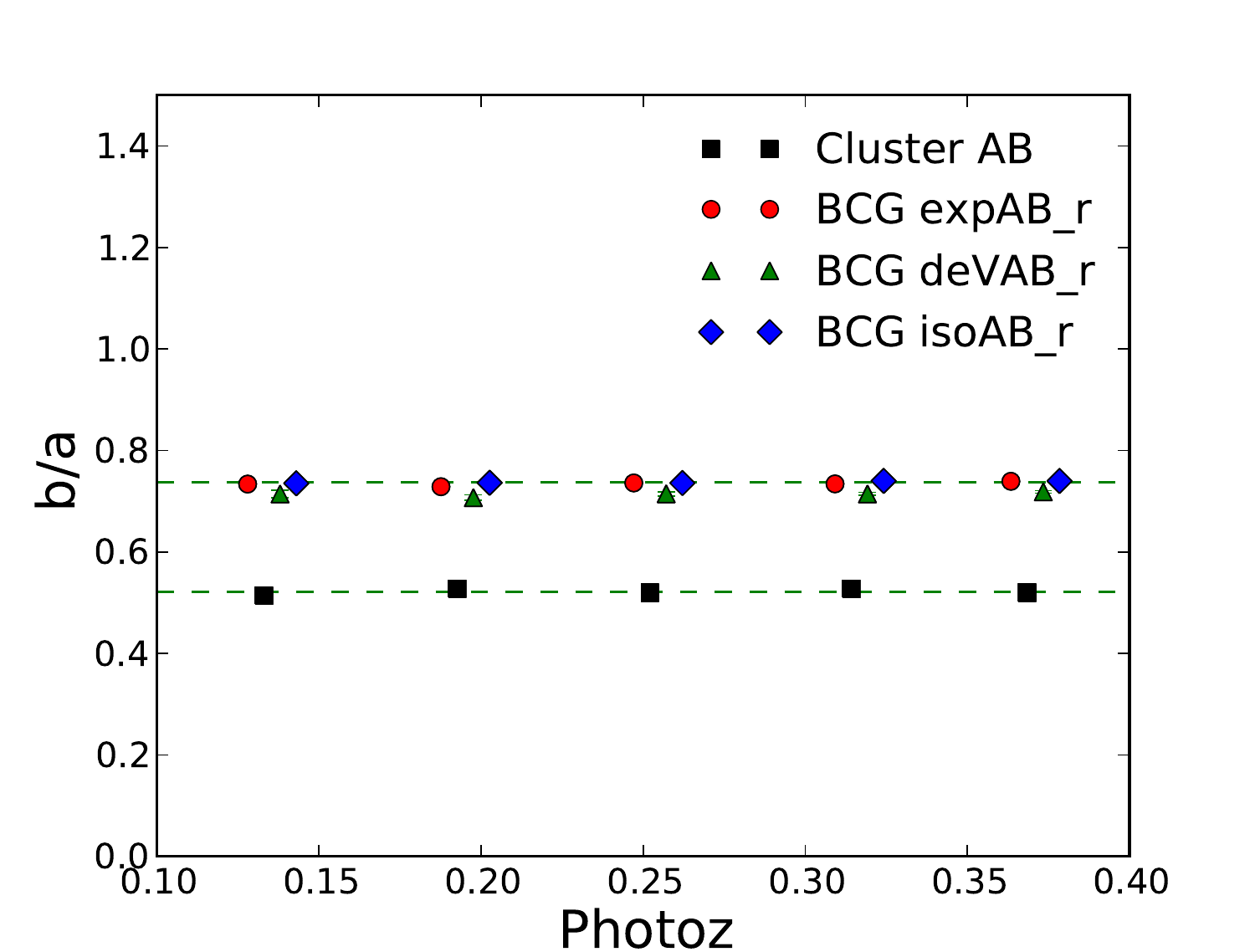}
\caption{The axis ratio $b/a$ of BCGs and clusters vs. redshift.}
\label{fig:ba}
\end{figure}

\begin{figure}
\epsscale{1}
\plotone{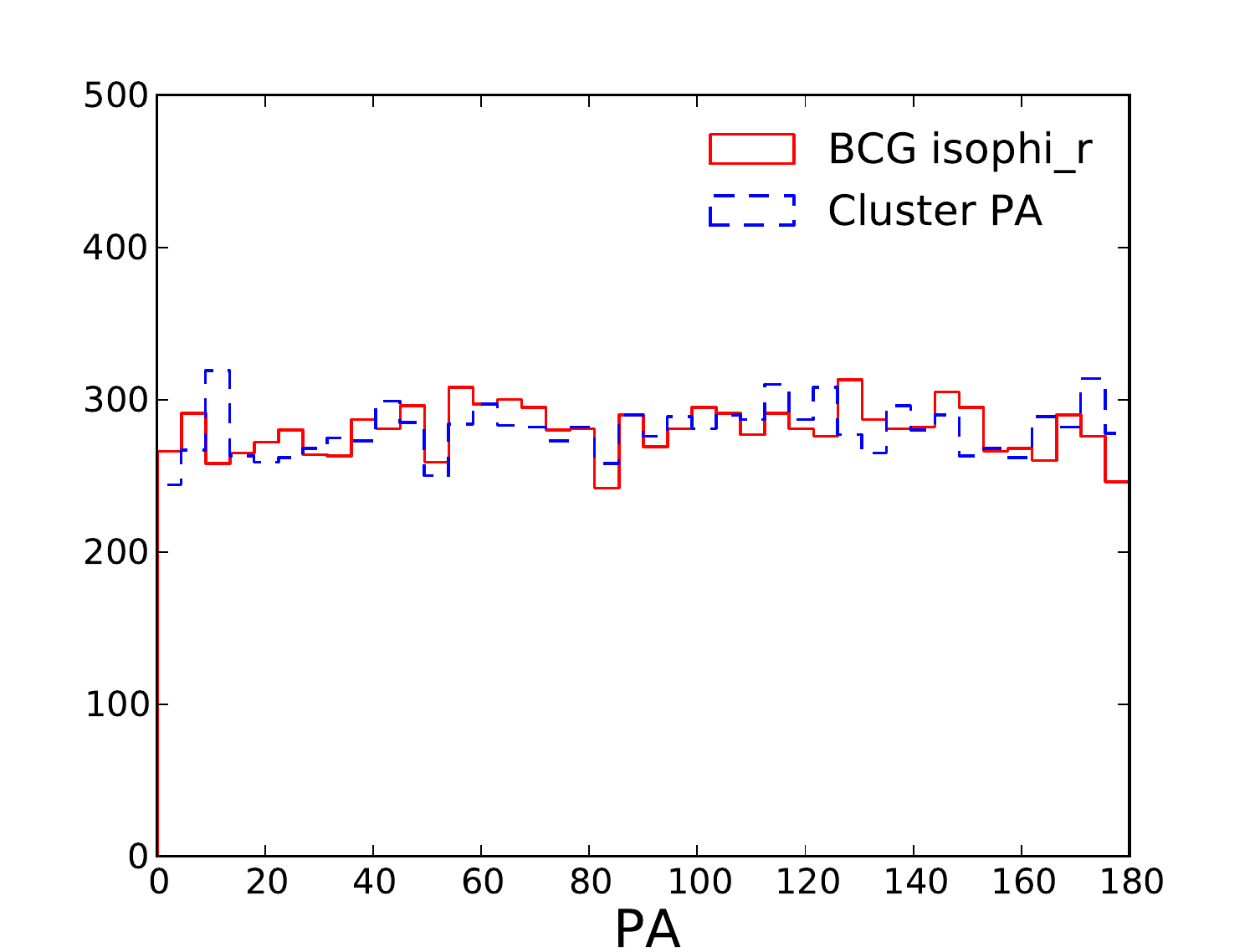}
\caption{The distribution of BCGs' and clusters' PAs.}
\label{fig:pa}
\end{figure}

To calculate the cluster $b/a$, we use the method as described in \citet{kim02,ostholt10}. For clusters, the axis ratio is defined as $b/a = (1-\sqrt{Q^2+U^2})/(1+\sqrt{Q^2+U^2})$, with the stokes parameters $Q=M_{xx}-M_{yy}$ and $U=2 M_{xy}$. The radius-weighted second moments are given by $M_{xx}=\left<x^2/r^2\right>$, $M_{yy}=\left<y^2/r^2\right>$ and $M_{xy}=\left<xy/r^2\right>$ with $r^2=x^2+y^2$. $x$ and $y$ are the distances between the satellite galaxies and BCG in the tangent plane. For BCGs, we use the measured $b/a$ in SDSS pipeline based on isophotal fit, exponential fit and DeVoucular fit. In Figure~\ref{fig:ba}, we plot the $b/a$ in each redshift bin of size 0.05. From the results, the axis ratio does not depend on redshift in a statistically significant way. Furthermore, we also checked that both the BCGs' PAs and cluster PAs are distributed randomly, as show in the right panel of Figure~\ref{fig:pa}. Therefore, the evolution of $\gamma$ in Figure~\ref{fig:gammaz} should not result from the S/N variation of the PA measurements.

\subsubsection{Magnitude Dependence}

\begin{figure}
\epsscale{1}
\plotone{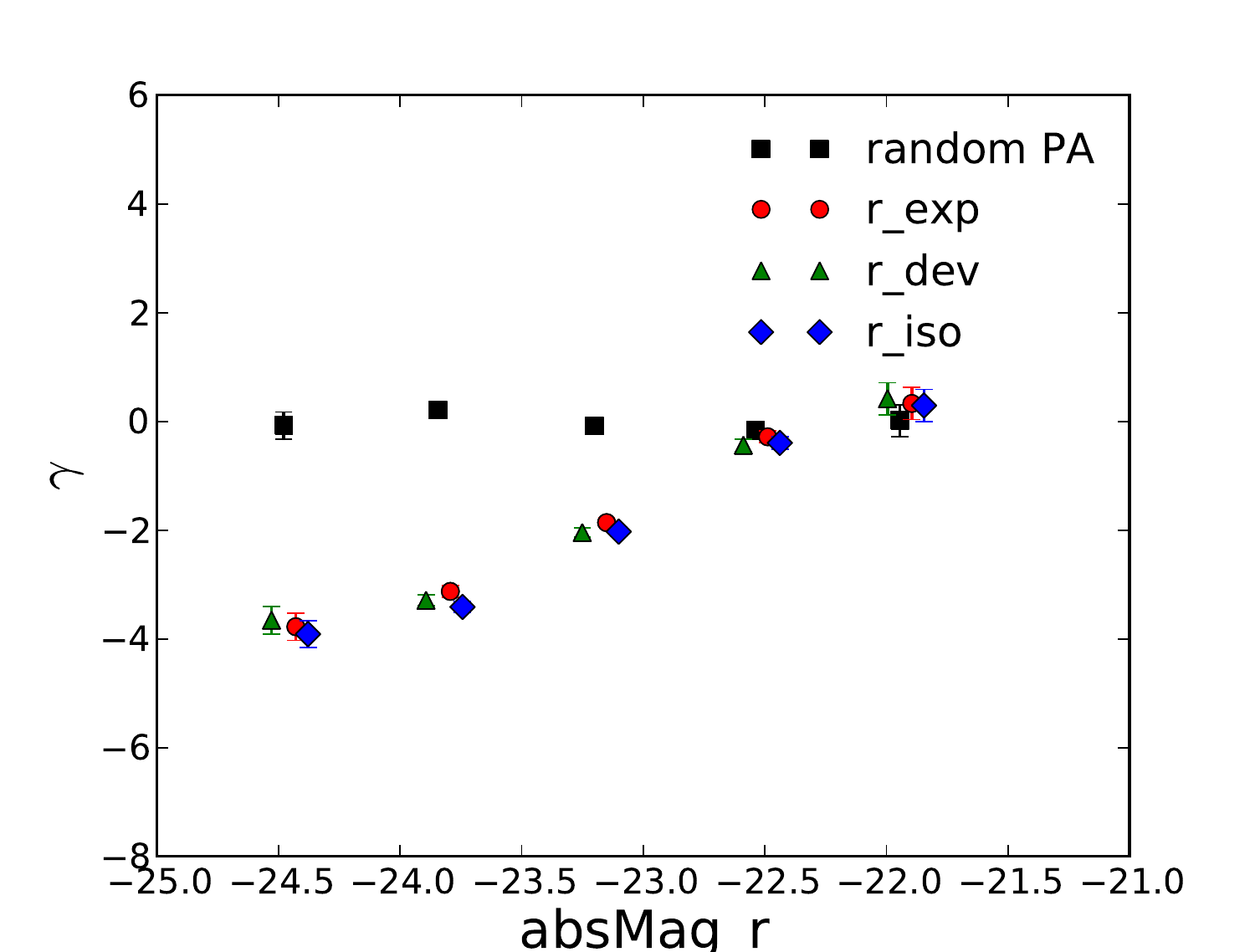}
\caption{The dependence of BCG alignment on the absolute $r$ band magnitudes of BCGs. }
\label{fig:gamma_ramag}
\end{figure}

We measure the BCG alignment vs the $r$ band absolute magnitudes of the BCG. The results are presented in Figure \ref{fig:gamma_ramag}

\begin{figure}
\epsscale{1}
\plotone{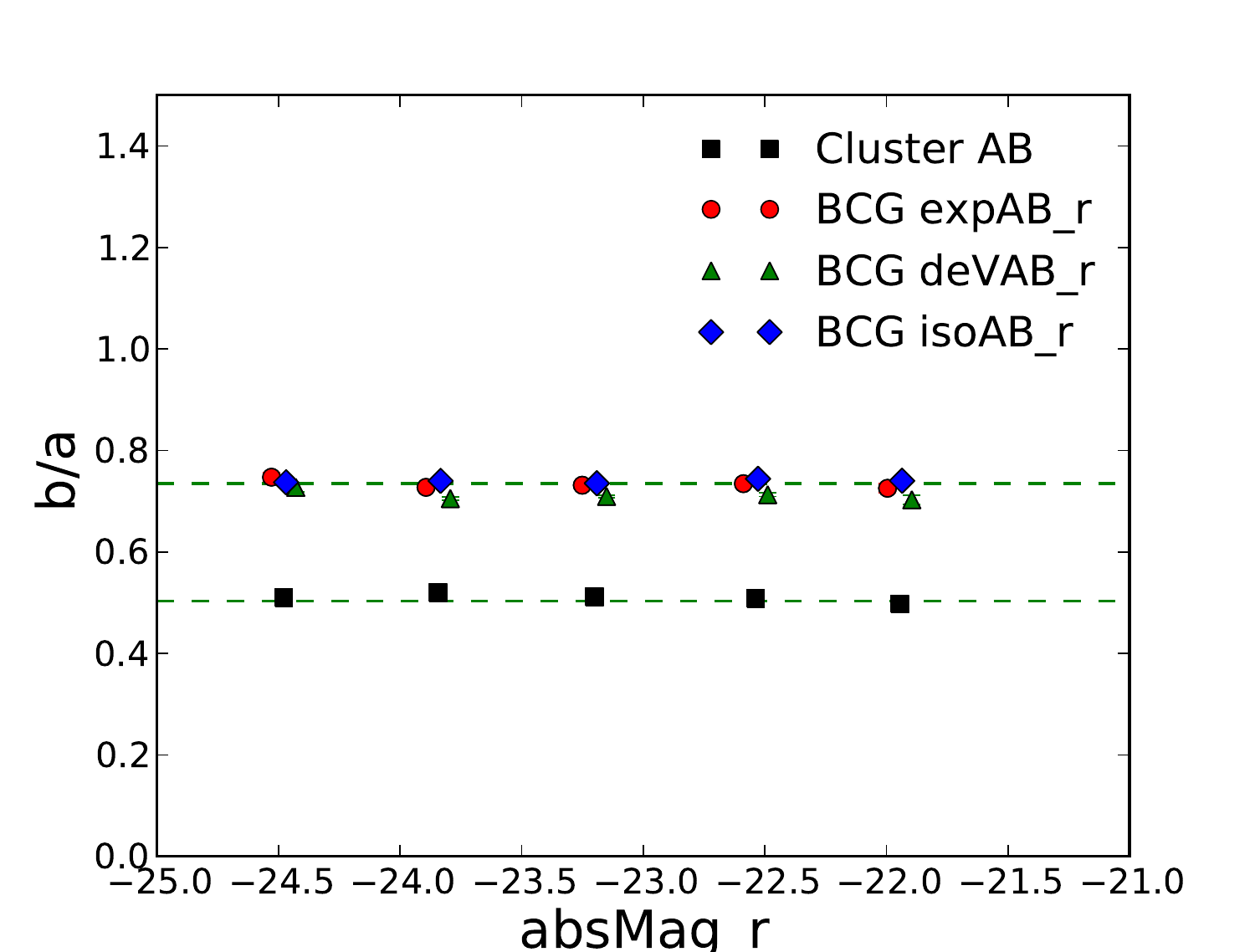}
\caption{The axis ratio $b/a$ of BCGs and clusters vs. the $r$ band absolute magnitude of BCG.}
\label{fig:ba_absmag}
\end{figure}

From the plot, we see that $\gamma$ strongly depends on the BCGs absolute magnitude. To further show that this is not due to S/N of the measurement of BCGs shape, we plot the axis ratio $b/a$ of BCGs and clusters as a function of the BCGs $r$ band absolute magnitude in Figure~\ref{fig:ba_absmag}. There is no dependence of $b/a$ on the BCGs absolute magnitude. 

\begin{figure*}
\begin{center}
\epsscale{1.2}
\plotone{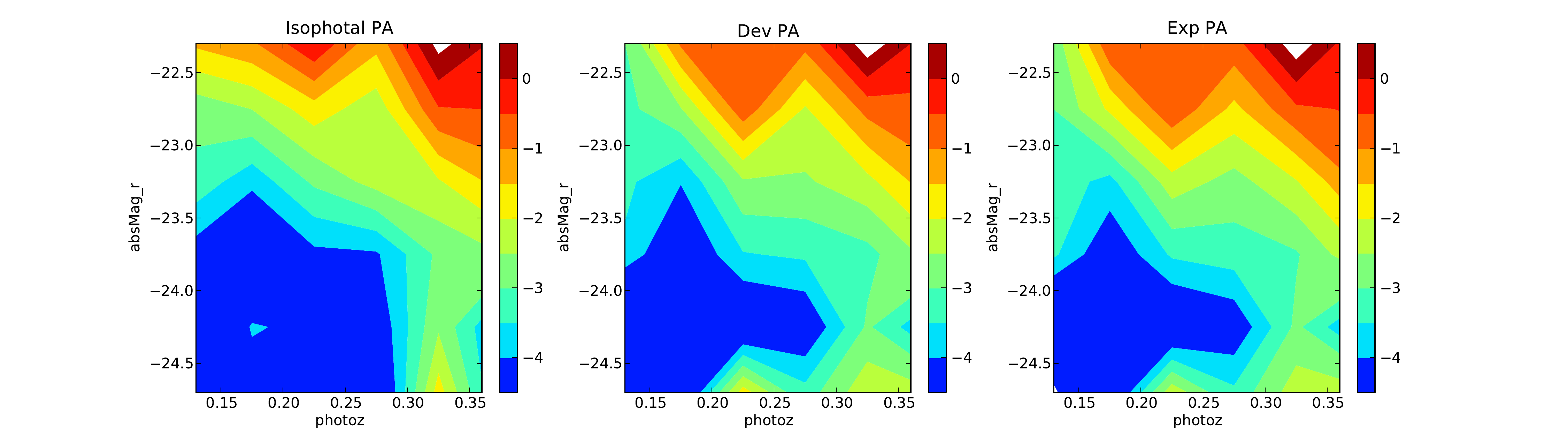}
\caption{Contours of $\gamma$ w.r.t. redshift and BCG absolute magnitude. The crossover of the contour lines are mainly due to the large error bar of each data point, which cannot be expressed in the contour plots.}
\label{fig:gamma_z_magbin}
\end{center}
\end{figure*}

From the above results, we see that the $\gamma$ depends on both redshift and the BCG absolute magnitude. To show this more clearly, we bin the cluster samples into photoz bins of size 0.05 and absolute magnitude bins of size 0.5. Then, we calculate the mean $\gamma$ in each bin and plot the results in Figure~\ref{fig:gamma_z_magbin}. The color in the plot indicates $\gamma$.

\begin{figure*}
\begin{center}
\epsscale{1.2}
\plotone{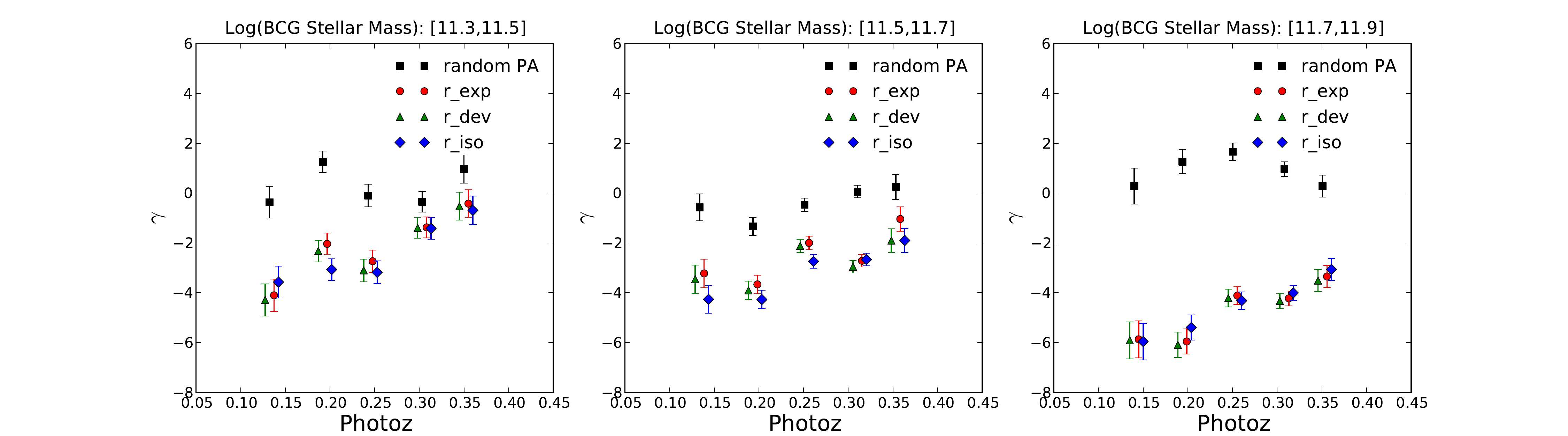}
\caption{Redshift dependence of $\gamma$ in three BCG stellar mass bins.}
\label{fig:gamma_z_stm}
\end{center}
\end{figure*}

From the plot, we can see that the $\gamma$ increases (i.e. absolute signal decreases) as redshift increases and as the absolute magnitude of the BCG decreases. Since the absolute magnitude of a galaxy is proportionally correlated with its mass, the above results indicate that the more massive BCGs tend to be more aligned with the cluster orientation. A subsample of the BCGs ($\sim$2800 BCGs) has their stellar masses measured in the MPA-JHU value-added catalog for SDSS DR7~\citep{kauffmann03,salim07}. This allows us to directly look at the trend of $\gamma$ with respect to the BCG stellar mass (a proxy of the total mass) and redshift. We choose three stellar mass bins and plot the $\gamma$ vs. redshift in each of them in Figure~\ref{fig:gamma_z_stm}. We can see that the $\gamma$ increases as redshift increases in each stellar mass bin and the massive BCGs shows more negative $\gamma$.



\subsubsection{Richness Dependence}
\begin{figure}
\epsscale{1}
\plotone{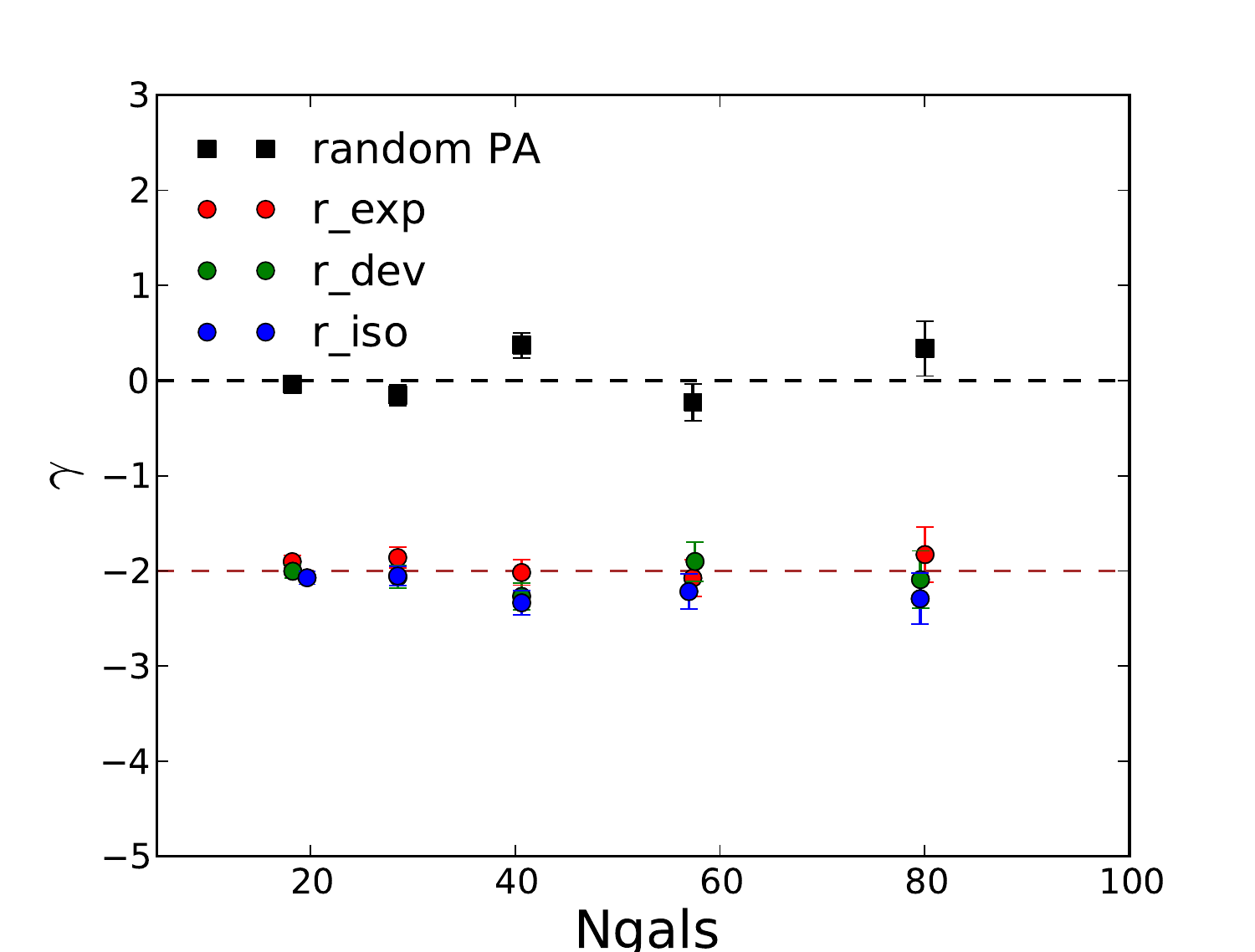}
\caption{Dependence of BCG alignment on cluster richness. Here, we bin the richness into bins with edges at 15, 25, 35, 50, 65 and 90. The results do not show a statistically significant dependence of BCG alignment on cluster richness.}
\label{fig:gammarich}
\end{figure}

Next, we check whether $\gamma$ depends on cluster richness. To do this, we bin the clusters by their richness in bins with edges at 15, 25, 35, 50, 65 and 90. We did not go to the richness bins above 90 due to the smaller number of clusters in that richness range. Then, we look at the mean $\gamma$ in each bin. Note that we did not re-bin them into different redshift bins to keep the number of clusters reasonably large. This will not affect our purpose for detecting richness dependence since clusters from different redshifts are randomly falling into different richness bins. We plot the results in Figure~\ref{fig:gammarich}. Based on the results, we do not see a dependence on cluster richness.

\subsection{Redshift Evolution of BCG Alignment Once Again}

Measuring the redshift evolution of BCG alignment is very important for understanding its origin. However, the redshift evolution of the measured alignment signal needs to be interpreted with great caution, especially for the cluster samples selected using photometric data. There are at least four factors that will introduce systematic redshift dependence and complicate the interpretation. (1) The S/N of the galaxy shape measurements will decrease as redshift increases. (2) When we look at clusters of different redshift, we need to make sure we are comparing the same population of satellite galaxies. That is, the cluster catalog needs to be volume limited in the redshift range. (3) The purity of clusters need to be consistently high across the redshift range. The change of purity will lead to decreased mean alignment signal. (4) The level of contamination from the projected field galaxies are prone to redshift dependence. Higher level of contamination will dilute the alignment signal. We have addressed (1) and (2) in previous sections, where we introduce the results. In the follows, we will focus on the (3) and (4). 

\begin{figure}
\begin{center}
\epsscale{1}
\plotone{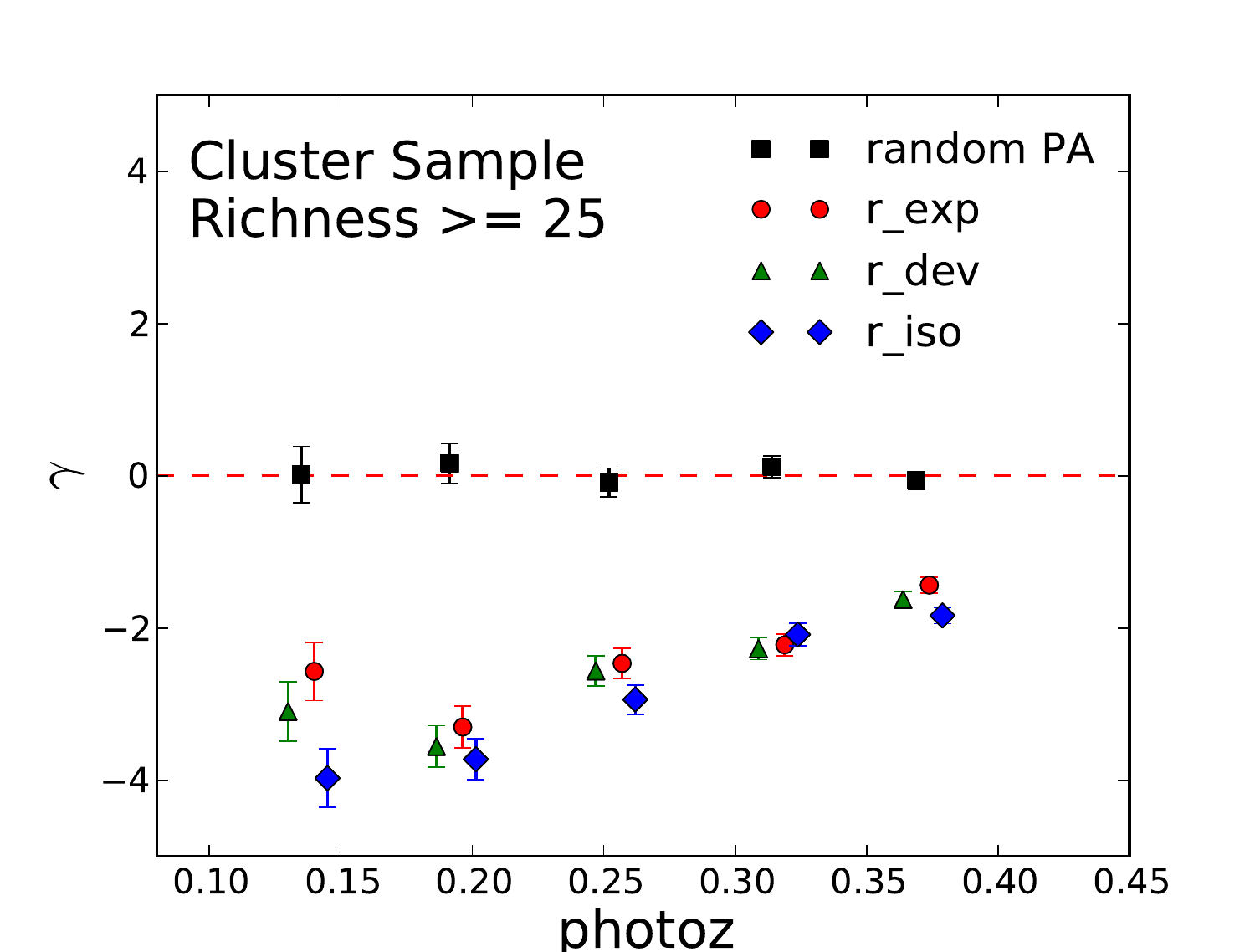}
\caption{BCG alignment at redshift bins of size 0.05 based on a subsample with higher purity and lower purity variations across the redshift range. The results indicate that the redshift dependence are not affected by the purity of the cluster sample we are using.}
\label{fig:gammaz25}
\end{center}
\end{figure}

The alignment signal from the falsely detected clusters should be consistent with zero, decreasing the mean alignment of the whole sample. For the subsample of GMBCG clusters with richness equal or greater than 15, it has been shown that the purity is consistently above 90\% and does not vary more than 10\% across the redshift range from 0.1 to 0.4~\citep{haocat}. To further show the purity variation will not produce the observed redshift dependence of $\gamma$, we choose another subsample of clusters with even higher purity. We choose clusters with richness greater than 25, which have a purity of above 95\% and vary less than 5\% in the redshift range. In Figure~\ref{fig:gammaz25}, we plot the BCG alignment parameter $\gamma$ from this subsample. Though the overall signal level increases a little at low redshift, the trend of redshift dependence does not differ much from the full sample with a lower richness threshold 15. So, the purity change should not explain the measured strong redshift dependence of $\gamma$. 

On the other hand, satellite galaxies selected using red sequence colors have different levels of contamination from projected field galaxies as redshift changes. This is mainly caused by the different degree of overlap between the red sequence population and the field galaxy population (e.g. see Figure 14 in \citet{haocat}). There are two competing effects that will increase or decrease the contamination. First, the separation between the red sequence component and the field galaxy component. As redshift increases, the two components separate farther, leading to decreased projection contamination in the red sequence. The second effect is the broadening of the distribution of both red sequence and field galaxy. As redshift increases, the measured width of red sequence increases mainly due to the photometric errors~\citep{haoecgmm}. This will increase the chance of projected field galaxies being identified as satellites when we select the satellite galaxies by color. Therefore, the actual contamination level is the compromise of these two effects. 

We can describe the measured alignment parameters, $\gamma$\footnote{Since we did not see a significant $\delta$ signal, we will consider only $\gamma$ in all the discussions hereafter. But the method described can also be applied to the $\delta$ case.}, as a combination of alignment from real cluster satellites and projected field galaxies. If we denote the alignment parameters from our measurements as $\gamma_m$, then we can decompose it into two parts as follows:

\begin{equation}\label{gamma_m}
\gamma_m = \frac{\sum_{i=0}^{N_c}\theta_i+\sum_{j=0}^{N_f}\theta_j}{N_c + N_f} - 45
\end{equation}

\noindent where $N_c$ is the number of true cluster satellite galaxies and $N_f$ is the number of projected field galaxies. We can introduce the fraction of real cluster satellite as $f_c(z) = N_c/(N_c + N_f)$, the BCG alignment from true cluster members as $\gamma_c=\sum_{i=0}^{N_c}\theta_i/N_c - 45$ and the BCG alignment from the projected field galaxies as $\gamma_f=\sum_{j=0}^{N_f}\theta_j/N_f - 45$. Substitute these definitions into Equation~\ref{gamma_m} and take ensemble average of the clusters, we will have:

\begin{equation}
\left<\gamma_m\right> = \left<f_c(z)\right> \left<\gamma_c\right> + \left[1 - \left<f_c(z)\right>\right] \left<\gamma_f\right>
\end{equation}

\noindent where $\left<...\right>$ denotes the average over the cluster ensemble. As the mean alignment signal from the field is consistent with zero, the alignment parameter $\gamma$ from the true cluster satellites is related to the measured one through the redshift dependent fraction $f_c(z)$. To the first order approximation, we can separate $f_c(z)$ into two parts as $f_c(z) = f_{const} \times f(z)$, where $f_{const}$ is a redshift independent component of the fraction, indicating the ``intrinsic'' fraction of true satellite based on color selection. $f(z)$ is the redshift dependent part, corresponding to the effect we described above. Then, the redshift dependence of the measured alignment $\gamma$ will be mainly determined by $f(z)$.  

\begin{figure}
\epsscale{1}
\plotone{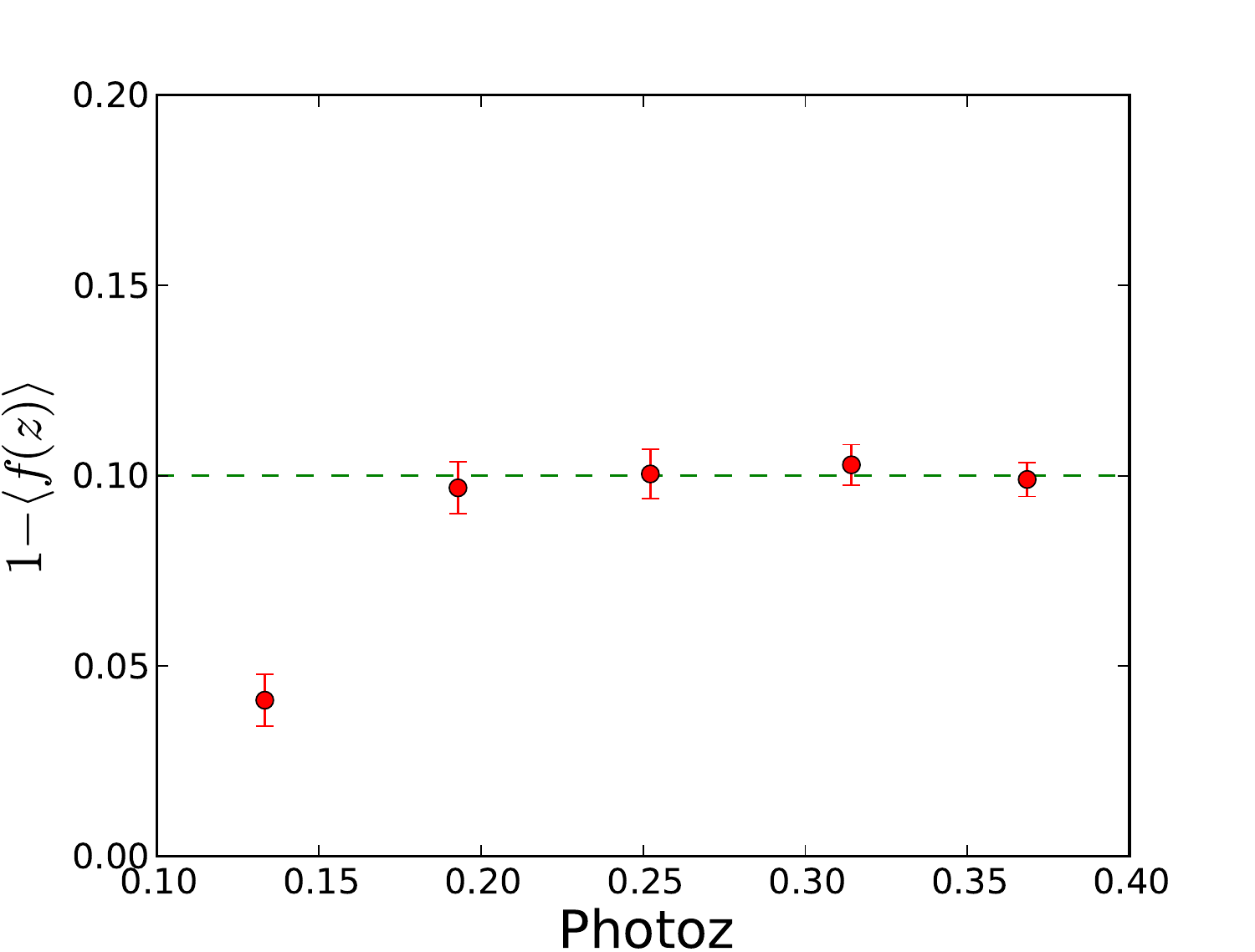}
\caption{Fraction of projected field galaxies at different redshift bin of size 0.05. It maintains constant except in the lowest redshift bin. }
\label{fig:fcz}
\end{figure}

In the GMBCG catalog, we also measured a weighted richness, which takes into account the different degree of overlaps between red sequence and the field galaxies at different redshift~\citep{haocat}. The difference between weighted richness and the direct member count richness is a good estimator of the number of projected galaxies due to the effect described above. The fraction of contamination can therefore be estimated by the ratio of this difference to the direct member count richness. In Figure~\ref{fig:fcz}, we plot the fraction of contamination ($1 - \left<f(z)\right>$) as a function of redshift in bins of size 0.05. The fraction is almost constant except for the lowest redshift bin. Again, this cannot explain away the dependence of $\gamma$ on redshift as shown in Figure~\ref{fig:gammaz} and Figure~\ref{fig:gammaz25}. Therefore, after considering all the possible systematics known to us, the measured redshift dependence of $\gamma$ still cannot be explained. In~\citet{ostholt10}, the authors also reported a different BCG alignment between one low redshift bin (0.08 - 0.26) and another high redshift bin (0.26 - 0.44), which is consistent with the results we find here. 

\subsection{Conclusions and Discussions}

We measure the satellite alignment and BCG alignment based on a large sample of photometrically selected galaxy clusters from the SDSS DR7. We detect a satellite alignment only when we use the isphotal PAs. As we noted in \S 3.3, the isophotal PA tends to trace the outer profile of the galaxy while the model fit PAs tend to trace the inner part of the galaxy. A direct interpretation of the measurement results could be that the outer part of the satellite galaxy is more susceptible to the the gravitational torque and thus shows an orientation preference toward the BCG. However the inner part of the galaxy is not affected much by the tidal torque and does not show preference toward the BCG. The measured discrepancy of the satellite alignment from different PAs could be a manifestation of the twisting of galaxy shape from inner part to outer part. However, another possibility of this discrepancy could be that the light from BCG contaminates the measurement of the PA based on the isophote fit to the outer region of the galaxy and lead to a ``artificial'' alignment. By comparing the dependence of $\delta$ on BCG apparent and absolute magnitudes, we favor the latter explanation. This means that, though the tidal torque within the galaxy cluster may induce the satellite alignment, we are not yet able to detect them based on our current SDSS data. It will be definitely an interesting question to address with the forthcoming high quality data such as that from the Dark Energy Survey~\citep{des05}.

For the BCG alignment, by introducing the alignment parameter $\gamma$, we detect a strong redshift and BCG absolute magnitude dependences of the alignment. The redshift dependence cannot be explained by our known systematics. This result implies that the BCGs orientation is a dynamically evolving process and gets stronger as the cluster system evolves. For the dependence of $\gamma$ on the absolute magnitude of BCG, our result is qualitatively consistent with the conclusion that clusters with BCG dominance show stronger BCG alignment in \citep{ostholt10}. Furthermore, based on a subsample of the BCGs whose stellar masses are available, we show that the BCG alignment signal becomes stronger as the BCG stellar mass increases. This result indicates that more massive BCGs (with lower absolute magnitude) are more likely to align with the major axes of clusters. 

We must take great caution when interpreting the dependence of $\gamma$ on BCG absolute magnitude and stellar mass since the purity of the cluster sample may also depend on the BCG absolute magnitude and stellar mass. As the cluster purity decreases, the alignment signal will decrease too. The faintest two bins in Figure~\ref{fig:gamma_ramag} show null alignment signal, which may also be due to the significantly decreased cluster purity. Nevertheless, we can still see a trend that $\gamma$ increases as the BCG absolute magnitude increases by looking at the bright end of the sample where we are confident about the cluster purity. Evaluating the cluster purity variation w.r.t BCG absolute magnitude turns out to be difficult because it requires a mock galaxy catalog that has BCG information properly built in. The way the mock catalog is constructed will impact the results significantly. Therefore, we think the best way to check this purity variation w.r.t. magnitude is to perform similar analysis with deeper data in the near future, such as the data from the upcoming Dark Energy Survey~\citep{des05}.

\bibliography{alg}

\section*{Acknowledgments}

JH thanks Scott Dodleson for helpful comments and Eric Switzer for helpful conversations. 
Funding for the SDSS and SDSS-II has been provided by the Alfred P.
Sloan Foundation, the Participating Institutions, the National
Science Foundation, the U.S. Department of Energy, the National
Aeronautics and Space Administration, the Japanese Monbukagakusho,
the Max Planck Society, and the Higher Education Funding Council for
England. The SDSS Web Site is http://www.sdss.org/.

The SDSS is managed by the Astrophysical Research Consortium for the
Participating Institutions. The Participating Institutions are the
American Museum of Natural History, Astrophysical Institute Potsdam,
University of Basel, University of Cambridge, Case Western Reserve
University, University of Chicago, Drexel University, Fermilab, the
Institute for Advanced Study, the Japan Participation Group, Johns
Hopkins University, the Joint Institute for Nuclear Astrophysics,
the Kavli Institute for Particle Astrophysics and Cosmology, the
Korean Scientist Group, the Chinese Academy of Sciences (LAMOST),
Los Alamos National Laboratory, the Max-Planck-Institute for
Astronomy (MPIA), the Max-Planck-Institute for Astrophysics (MPA),
New Mexico State University, Ohio State University, University of
Pittsburgh, University of Portsmouth, Princeton University, the
United States Naval Observatory, and the University of Washington.

\label{lastpage}

\end{document}